\documentclass{JHEP3}
\usepackage{latexsym,bm,amsmath,amssymb,amsfonts, dsfont}
\usepackage{epsfig,graphics,graphicx,mathrsfs}
\usepackage{slashed}
\usepackage{cite}
\usepackage[latin1]{inputenc}

\def\xb{{\bf x}}

\def\rb{{\bf r}}    
\def\bb{{\bf b}}

\newcommand{\beeq}{\begin{eqnarray}}
\newcommand{\eeeq}{\end{eqnarray}}
\newcommand{\be}{\begin{equation}}
\newcommand{\ee}{\end{equation}}
\newcommand{\bea}{\begin{array}}
\newcommand{\eea}{\end{array}}

\def\be{\begin{equation}}
\def\ee{\end{equation}}
\def\bea{\begin{eqnarray}}
\def\eea{\end{eqnarray}}

\title{\rm   Exclusive vector meson production and small-x evolution}
\author{J.~Berger$^{a}$ and A.M.~Sta\'sto$^{a,b,c}$ \\
\!\!$^{a}$The Pennsylvania State University, University Park, PA 16802,  USA\\
\!\!$^{b}$RIKEN Center, Brookhaven National Laboratory, Upton, NY 11793, USA \\
\!\!$^{c}$H.~Niewodnicza\'nski Institute of Nuclear Physics, Polish Academy of Sciences, Cracow, Poland \\
\\
E-mail:  \email{jxb1024@psu.edu, astasto@phys.psu.edu}}

\abstract{
The process of exclusive elastic vector meson production in deep inelastic scattering is investigated within  the dipole model framework supplemented by the small $x$ evolution. 
The dipole-proton amplitude is obtained from the nonlinear Balitsky-Kovchegov evolution equation with impact parameter dependence.  This dipole amplitude is used to compute the differential cross section for exclusive production of J/$\Psi$, $\phi$, and $\rho$ vector mesons. These numerical calculations are compared with the wide range of experimental data from HERA.  Good agreement between the experimental data and the calculations is found.}
\begin{document}

%%%%%%%%%%%%%%%%%%%%%%%%%%%%%%%%%%%%%%%%%
\section{Introduction}
\label{sec:intro}

Exclusive diffractive vector meson production in deep inelastic scattering,
 $\gamma^*p\rightarrow Vp$,  has been extensively studied at high energies at HERA electron(positron)-proton collider.
Both H1 \cite{Adloff:1999kg,Aktas:2005xu,Aaron:2010} and ZEUS \cite{Chekanov:2002xi,Chekanov:2004mw,Chekanov:2005cqa,Chekanov:2007zr} collaborations   performed detailed measurements of this process 
in a wide kinematic range. In this process  the final state consists of the elastically scattered proton accompanied by  a vector meson. It provides important information about the electron-proton interaction at small $x$ as well as  the size of the proton in impact parameter. It  can be also used to constrain the form of the vector meson wave functions. The main conclusions from these measurements can be summarized as follows (for concise reviews of the experimental  results see for example \cite{Levy:2007fb,Bunyatyan:2008zza}).  The cross section for the elastic vector meson production increases with the energy for all values of the (minus) photon virtuality $Q^2$.  This dependence can be phenomenologically parametrized as $\sim W^\delta$ where $W$ is the energy of the virtual photon-proton system in DIS. The exponent $\delta$
depends strongly on the scales involved in the process, namely on the $Q^2$ and the mass $M_V$  of the vector meson.  The exponent $\delta$ shows a marked increase with both $M_V$ and $Q^2$ scales. For example, for $\rho$ production the variation of $\delta$ with $Q^2$ is from $0.2$ for photoproduction up to $\sim 0.8$ for values of $Q^2\sim 35 \;{\rm GeV}^2$. This is  consistent with the picture of the soft-hard transition from small  to large scales. The value of the $\delta$ in the  photoproduction region,  and at low $Q^2$, is consistent with the  soft exchange,  while in the high $Q^2$ region perturbative gluon exchange is the dominating process. For low values of $Q^2$ multiple interactions between the quark-antiquark system and a proton also lead to smaller values of $\delta$, see for example \cite{Kowalski:2006hc}. For $J/\Psi$ production the variation of $\delta$ with $Q^2$ is negligible in the measured range (from photoproduction to electroproduction with $Q^2 \sim 30 \; {\rm GeV}^2$), and its value is about $\sim 0.8$, indicating that the heavy mass of the $J/\Psi$ vector meson provides the hard scale in this process.

The measurement of the   momentum transfer $t$ in this process provides the information about the transverse size of the interaction region, see for example \cite{Frankfurt:2010ea}. These measurements indicate a strong dependence of the differential cross section on $t$, which can be parametrized by an exponential $\exp(-B_D |t|)$.
The experimental data show that the value of $B_D$ decreases from 
about $\sim10-12 \; {\rm GeV}^{-2}$ at lowest values of $(Q^2+M_V^2)$ to about $5 \; {\rm GeV}^{-2}$ at scale $Q^2+M_V^2\sim 10 \; {\rm GeV}^2$, and stays constant for larger scales.  This can be interpreted as follows: the value of parameter $B_D$ is closely related to the transverse size of the interaction region which is a combination (or more precisely, a convolution) of the sizes set by the vector meson and the proton. At low values of $Q^2+M_V^2$ it is the first size that prevails, and as a result the value of  $B_D$ strongly varies with this scale. At larger values of $Q^2+M_V^2$ the data indicate that the vector meson size is much smaller than the typical size of the proton, (or more precisely the radius of the gluon density in the proton), and it is this 'gluonic' size of the proton which is dominating $B_D$ and thus visible at larger values of $Q^2+M_V^2$. This value is about $5\;{\rm GeV}^{-2}$ or equivalently $\sim 0.6 \;{\rm fm}$, and  it is   smaller than the electromagnetic size of the proton. The data also indicate a slow variation of the slope $B_D$ with the increasing energy of the interaction $W$. The slope can be parametrized by the following Regge inspired formula
$B_D=B_0+4\alpha'_{I\!P}\ln (W/W_0)$, with the fitted value of $\alpha'_{I\!P}\sim 0.115\pm0.018 (+ 0.008)(-0.015)$ \cite{Chekanov:2002xi} and $\alpha'_{I\!P}\sim0.164\pm0.028\pm0.030$ \cite{Aktas:2005xu} for $J/\Psi$ photoproduction  (parameter $W_0=90\;{\rm GeV}$).
This shows that the size of the gluon distribution in impact parameter space is increasing with rising energy, which is indicative of Gribov-type diffusion of gluons in transverse space.

In this paper we analyze elastic diffractive vector meson production using the dipole model approach for small $x$ physics. The novel aspect of our analysis is that the dipole scattering amplitude is obtained from the numerical solution to the impact parameter dependent Balitsky-Kovchegov (BK) equation \cite{Balitsky:1995ub,Kovchegov:1999yj,Kovchegov:1999ua}.
Numerous studies, which utilized the dipole model, were performed up to date 
whose goal was a description of the exclusive diffractive vector meson production at small $x$, see for example \cite{Nemchik:1994fp,Frankfurt:1995jw,Nemchik:1996cw,Munier:2001nr,Forshaw:2003ki,Kowalski:2006hc,Flensburg:2008ag}.
As mentioned above, the $t$-dependence  of the vector meson production cross section is particularly interesting as it is directly related to the transverse size of the interaction region. In the framework of the dipole model this is encoded in the dipole-proton scattering amplitude and its dependence on the impact parameter.  The increase  of the $t$ slope, $B_D$, with the energy $W$ should be directly coupled with the increase of the transverse interaction area caused by the diffusion of the dipoles in coordinate space. 
This diffusion in transverse coordinate space is  present in the dipole branching Monte-Carlo \cite{Avsar:2005iz,Avsar:2006jy,Avsar:2007xh,Flensburg:2008ag} as well as the Balitsky-Kovchegov equation when   the impact parameter dependence is taken into account \cite{GolecBiernat:2003ym,Berger:2010sh}.  The main goal of the present work is to use the numerical  solution to the BK equation, which includes the full impact parameter dependence, and verify its compatibility with the experimental data on the exclusive vector meson production in deep inelastic lepton-proton scattering. 

The outline of the paper is the following, in the next section we recall the formalism of exclusive vector meson production in the dipole model, in Subsec.~\ref{subsec:dipscat} we discuss the basic facts which pertain to the solution of the BK equation with impact parameter dependence, and discuss modifications due to confinement. In Subsec.~\ref{subsec:phenom}
additional modifications are outlined, such as the non-perturbative modification of the photon wave function and inclusion of the skewed effect into the initial gluon distribution, as well as inclusion of the real part of the scattering amplitude. This is followed by the description of the model for the vector meson wave function which is used for the calculation. In Sec.~\ref{sec:results}
we present our results, which include  the comparison of the theoretical calculation with a wide range of experimental data from HERA on $\rho,\phi$ and $J/\Psi$ production. We compute the cross section integrated over the momentum transfer $t$  and investigate its $W$ and $Q^2$ dependence. Then  we compute the differential cross section with $t$ dependence in bins of $W$ and $Q^2$.
The experimentally measured  scale and energy dependence of the slope $B_D$ is also compared with theoretical calculations. Finally, in the last section we state  the conclusions and present an outlook.

%%%%%%%%%%%%%%%%%%%%%%%%%%%%%%%%%%%%%%%%%
\section{Exclusive vector meson production  in the dipole model}
\label{sec:CSdipole}

\begin{figure}
\begin{center}
\includegraphics[width=0.45\textwidth]{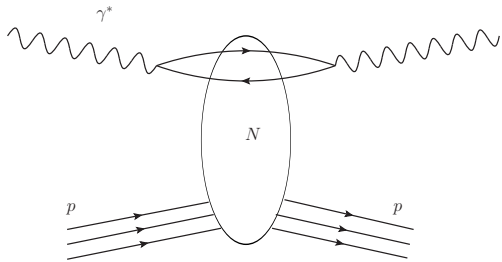}
\caption{Schematic representation of the inclusive cross section for deep inelastic scattering in the dipole model. Virtual photon fluctuates into $q\bar{q}$ pair and interacts with the hadronic target. The dynamics of the interaction is encoded in the dipole - target scattering amplitude $N$.}
\label{fig:dipolemodel}
\end{center}
\end{figure}

The dipole model \cite{Nikolaev:1990ja,Nikolaev:1991et} is a very useful tool in evaluating many processes at small values of $x$.  One of the advantages of this approach is the possibility of including the multiple parton scattering effects. It has been originally formulated for the description of deep inelastic lepton-proton (or nucleus) scattering at small $x$. In this picture, utilizing the leading logarithmic approximation in $x$, the incoming electron emits a virtual photon which fluctuates into a quark-antiquark pair (a dipole).  This color dipole then subsequently interacts with the parton constituents of the nucleon, as  is illustrated in Fig.~\ref{fig:dipolemodel}. 

%%%%%%%%%%%%%%%%%%%%%%%%%%%%%%%%%%%%%%%%%
%%%%%%%%%%%%%%%%%%%%%%%%%%%%%%%%%%%%%%%%%

The interaction of the dipole pair with the target is given by the scattering amplitude $N$. This quantity is non-perturbative in principle, however its small $x$ evolution can be found from the BK equation.  The $q\bar{q}$ pair is characterized by a dipole size which is defined as a separation distance of the color charges $\xb_{01} = \xb_0 - \xb_1$ (where $\xb_0$ and $\xb_1$ are the positions of the $q$ and $\bar{q}$ in transverse space).\footnote{  In this paper we shall denote vector quantities in bold, otherwise they should be read as magnitudes of the associated vector.  Also, alternatively we will be also using here the notation for the dipole size  to be  $r=x_{01}$ and impact parameter $b=\frac{|\xb_0 + \xb_1|}{2}$. } The transverse momentum of the quarks in the dipole  is on the order of $\sim \frac{1}{x_{01}}$ where large dipoles correspond to the infra-red region and need to be regulated.  The scattering amplitude $N(\rb,\bb;Y)$  contains all the information about the dynamics of the strong interaction. It depends on the dipole size $\rb$, on the impact parameter of the dipole with respect to the target $\bb$ and the rapidity $Y$. In the following analysis the   full dependence of the scattering amplitude on the impact parameter $\bb$  will be taken into account.  

The 
 structure functions $F_2$ and $F_L$ for the proton can be evaluated using the following standard formulae in the dipole picture in the transverse coordinate representation

\be
F_2(Q^2,x) = \frac{Q^2}{4 \pi^2 \alpha_{em}}\int{d^2 {\bf r} \int_0^1 dz \left(|\Psi_T(r,z,Q^2)|^2+|\Psi_L(r,z,Q^2)|^2\right) \sigma_{\rm dip}({\bf r},x)} \; ,
\label{eq:F2}
\ee
and
\be
F_L(Q^2,x) = \frac{Q^2}{4 \pi^2 \alpha_{em}}\int{d^2 {\bf r} \int_0^1 dz |\Psi_L(r,z,Q^2)|^2 \sigma_{\rm dip} (\rb,x)} \; .
\label{eq:FL}
\ee
The  dipole-proton cross section $\sigma_{\rm dip}$ can be obtained from the  scattering amplitude by integrating over the impact parameter $\bb$
\be
\sigma_{\rm dip}(\rb,x) = 2 \int{d^2\bb \, N(\rb,\bb;Y)} \; ,  \; \; \; \; \; \;Y=\ln 1/x \; .
\label{eq:sigmadip}
\ee
Since the amplitude $N$ is dimensionless and the integration is over the impact parameter, the dipole cross section $\sigma_{\rm dip}$ has obviously a dimension of the area. We see therefore that although the inclusive quantities are sensitive to the size of the interaction area, the details of the impact parameter profile are not directly accessible through this process.

The quantities  $\Psi(\rb,Q^2,Y)_{T/L}$  are the  photon wave functions. They describe the   dissociation of a photon into a $q$$\bar{q}$ pair and can be calculated from perturbation theory. The photon wave function has the following form for the case of transverse photon polarization
\be
|\Psi_T(r,z,Q^2)|^2 = \frac{3 \alpha_{em}}{2 \pi^2} \sum_f e_f^2\left(\left[z^2 + (1-z)^2\right]\bar{Q}^2_f K_1^2\left(\bar{Q}_f r\right) + m_f^2 K_0^2\left(\bar{Q}_f r\right)\right) \; ,
\label{eq:PhotonT}
\ee
and for longitudinal polarization
\be
|\Psi_L(r,z,Q^2)|^2 = \frac{3 \alpha_{em}}{2 \pi^2} \sum_f e_f^2\left(4Q^2z^2(1-z)^2K_0^2\left(\bar{Q}_f r\right)\right) \; .
\label{eq:PhotonL}
\ee
In the above equations  $\bar{Q}^2_f = z(1-z)Q^2 + m_f^2$, where $-Q^2$ is the photon virtuality and $z,(1-z)$ are the fractions of the longitudinal momentum of the photon carried by the quarks. In addition  $K_{0,1}$ are modified Bessel functions of the second kind.  The summations are over the active quark flavors $f$, of charge $e_f$, and mass $m_f$.  

%%%%%%%%%%%%%%%%%%%%%%%%%%%%%%%%%%%%%%%%%
\begin{figure}
\begin{center}
\includegraphics[width=0.45\textwidth]{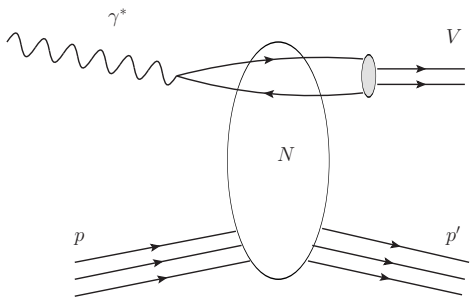}
\caption{Schematic representation of exclusive vector meson production in the dipole model at  small $x$. A virtual photon fluctuates into a $q\bar{q}$ pair and interacts with the target(proton). After the interaction the vector meson $V$ is formed which is measured in the final state. The proton scatters elastically with some momentum transfer $t$.}
\label{fig:dipolemodel2}
\end{center}
\end{figure}
%%%%%%%%%%%%%%%%%%%%%%%%%%%%%%%%%%%%%%%%%

The dipole picture  can be also used to compute diffractive processes. Here, we are interested in the process of the exclusive, diffractive production of the vector-meson $\gamma^* p \rightarrow V p'$. The amplitude for this  process is schematically illustrated in Fig.~\ref{fig:dipolemodel2}.  The virtual photon still fluctuates into $q\bar{q}$ pair, which then interacts with the proton and a vector meson is formed, which is  measured in the final state. The proton scatters elastically, its 4-momentum in the initial state is $p$ and in the final state is $p'$.
The formula for the amplitude for this process reads

\be
A(x,\Delta,Q) \; = \; \sum_{h,\bar{h}} \, \int d^2 \rb \int dz \, \Psi_{h,h^*}(\rb,z,Q^2) \, {\cal N}(x,\rb,\Delta) \,  \Psi_{h,h^*}^V(\rb,z)  \; ,
\label{eq:amplitude}
\ee
where $h$($\bar{h}$) is the helicity of quark (antiquark) and  $\Psi_{h,\bar{h}}^V$ is the vector meson wave function. $\Delta$ is the 2-dimensional momentum transfer related to the Mandelstam variable $t=-\Delta^2$.
 
The differential cross section for the process is given by
 \be
 \frac{d\sigma}{dt} \; = \; \frac{1}{16 \pi} |A(x,\Delta,Q)|^2 \; .
 \ee
 The amplitude ${\cal N}(x,\rb,\Delta)$ can be related to the scattering amplitude $N(x,\rb,\bb)$ introduced earlier, the amplitude in the impact parameter representation through the appropriate 2-dimensional Fourier transform
 
 \be
 {\cal N}(x,\rb,\Delta) = 2 \int d^2 \bb \, N(x,\rb,\bb) \, e^{i \Delta \cdot \bb} \; .
 \label{eq:Fourier}
 \ee
 In this notation the dipole  cross section, (compare (\ref{eq:sigmadip})), is 
 \be
 \sigma_{\rm dip}(x,\rb) \; = {\rm Im} \,i{\cal N}(x,\rb,\Delta=0) \;,
 \ee
 which is the expression for the optical theorem for scattering of dipoles.
 
 This process, through its dependence on the momentum transfer $t$, offers a unique possibility of constraining the impact parameter profile of the dipole scattering amplitude.
 
 Formulae \eqref{eq:amplitude} and \eqref{eq:Fourier} were original expressions derived under the assumption
 that the dipole size is much smaller than the proton. In Ref.~\cite{Bartels:2003yj}, a correction due to the finite size 
 of the dipole was calculated. It was shown that in the non-forward case, $\Delta\neq 0$, the amplitude can be
 written in the similar form as above with the modification of the \eqref{eq:Fourier} to include the exponential factor
 $\exp(-i(1-z)\rb \cdot \Delta)$ in the following way
  \be
 {\cal N}(x,\rb,\Delta,z) = 2 \int d^2 \bb \, N(x,\rb,\bb) \, e^{i \Delta \cdot (\bb-(1-z)\rb )} \; .
 \label{eq:Fouriermod}
 \ee
 This modification was included in the calculation \cite{Kowalski:2003hm}
 and it was shown that it  has a non-negligible effect on  cross sections, especially on the values of the $B_D$ slope which controls the $t$-dependence as a function of the scale $Q^2+M_V^2$.
 
%%%%%%%%%%%%%%%%%%%%%%%%%%%%%%%%%%%%%%%%%
\subsection{Dipole scattering amplitude from impact parameter dependent BK evolution}
\label{subsec:dipscat}

The dipole-proton scattering amplitude $N(\rb,\bb;Y)$ at high values of rapidity $Y$ (or small $x$) is found from the solution to the nonlinear integro-differential Balitsky-Kovchegov (BK) evolution equation \cite{Balitsky:1995ub,Balitsky:1998ya,Kovchegov:1999yj,Kovchegov:1999ua}. The BK evolution equation can be represented in the following form:
\be
\frac{\partial N_{\xb_0\xb_1}}{\partial Y} = \int\frac{d^2\xb_2}{2\pi}\, {\cal  K}(x_{01},x_{12},x_{02}; \alpha_s,m)\left[N_{\xb_0\xb_2}+N_{\xb_2\xb_1}-N_{\xb_0\xb_1}-N_{\xb_0\xb_2}N_{\xb_2\xb_1}\right] \; .
\label{eq:BK}
\ee

In the above equation we used the shorthand notation for the arguments of the amplitude $N_{\xb_i\xb_j}\equiv N({\rb_{ij}=\xb_i-\xb_j},\bb_{ij}=\frac{1}{2}(\xb_i+\xb_j);Y)$
which  depends on the two transverse positions $\xb_i$ and $\xb_j$ and on the rapidity $Y$.  The branching kernel ${\cal K}(x_{01},x_{12},x_{02}; \alpha_s,m)$ depends on the dipole sizes involved and contains all information about the splitting of the dipoles. In addition, it depends on  the running coupling $\alpha_s$. The way the strong coupling runs will be specified later in this work. We have also indicated that the kernel depends  on the infra-red cutoff $m$ which we impose  in order to regulate large  dipoles.  

 Eq.~(\ref{eq:BK}) is a differential equation in rapidity and hence suitable initial conditions need to be specified at some initial value of rapidity $Y=Y_0$.
 As in the previous work \cite{Berger:2011ew} we are choosing to use the initial condition in the form of the Glauber - Mueller parametrization with (most of) the parameters equivalent to those used in Ref.~\cite{Kowalski:2003hm}

\be
N_{\rm GM}(r,b;Y=\ln 1/x) \, = \, 1 - \exp{\left(-\frac{\pi^2}{2N_c}r ^2 x g(x,\eta^2) T(b)\right)} \; ,
\label{eq:glaubermueller}
\ee
with 
\be
T(b) \, = \, \frac{1}{8 \pi} e^{\frac{-b^2}{2B_G}} \; .
\label{eq:profile} 
\ee

 In formula (\ref{eq:glaubermueller}) the function $xg(x,\eta^2)$ is the integrated gluon density function and $T(b)$ is the density profile of the target in transverse space with the extension set by the parameter $B_G$.  
The  integrated gluon density in (\ref{eq:glaubermueller}) was also taken from fits performed in \cite{Kowalski:2006hc}. 
Scale parameter  in the gluon density is set to be $\eta^2=\mu_0^2+\frac{C^2}{r^2}$ with parameters $\mu_0$ and $C=2$ set to obtain the best description of the data. The values of these parameters are given in Table~\ref{Table:Parameters}.
We use (\ref{eq:glaubermueller}) as the initial condition at $Y_0=\ln 1/x_0$, $x_0=10^{-2}$ and evolve the amplitude with the BK equation to obtain the solution at lower values of $x<x_0$. 
We also note that the initial condition (\ref{eq:glaubermueller}) depends only on the absolute values of the dipole size and impact parameter. A nontrivial dependence on the angle between vectors $\rb$ and $\bb$ is not present in the initial condition, instead being dynamically generated when the initial condition is evolved with the BK equation.

The BK equation was solved numerically by discretizing the scattering amplitude in terms of  variables  $(\log_{10}r,\log_{10}b,\cos \theta)$, where $\theta$ is the angle between the impact parameter $\bb$ and the dipole size $\rb$.    The amplitude $N(r,b,\cos\theta)$ was placed on a grid with dimensions $200_r\times200_b\times20_\theta$. More details about the techniques and the properties of this solution can be found in Refs.~\cite{GolecBiernat:2003ym,Berger:2010sh}.

The running of the coupling in the BK kernel is a next-to-leading effect which has been evaluated in \cite{Balitsky:2006wa,Kovchegov:2006vj}.  In this calculation we utilize the  prescription from Ref.~\cite{Balitsky:2006wa} which is of the form
\begin{equation}
{\cal K} =  \bar{\alpha}_s(x_{01}^2)\left[\frac{1}{x_{02}^2}\left(\frac{\alpha_s(x_{02}^2)}{\alpha_s(x_{12}^2)} - 1\right) + \frac{1}{x_{12}^2}\left(\frac{\alpha_s(x_{12}^2)}{\alpha_s(x_{02}^2)} - 1\right) + \frac{x_{01}^2}{x^2_{12} x_{02}^2}\right] \;.
\label{eq:KernLOBal}
\end{equation}
In the above equation, we use 

\be
\alpha_s(x^2) = \frac{1}{b \, \ln\left[\Lambda_{\rm QCD}^{-2}\left(\frac{1}{x^2} + \mu^2\right)\right]}  \; ,
\label{eq:coupling}
\ee

Here $b = \frac{33 - 2n_f}{12\pi}$ and $n_f$ is the number of active flavors. The $\mu$ parameter effectively freezes the coupling at  large dipole sizes at $\alpha_{s,{\rm fr }} = \frac{1}{b \ln\left[\Lambda^{-2}\mu^2\right]}$. In our simulations we used  $\mu=0.52 \; {\rm GeV}$ as an infra-red regulator for the strong coupling, $\Lambda_{\rm QCD}=0.246 \; {\rm GeV}$, and the rescaled strong coupling is defined as  $\bar{\alpha}_s=\frac{\alpha_s N_c}{\pi}$.

The evolution  equation  \eqref{eq:BK}  has been derived in perturbation theory and it does not include any effects of confinement. The branching kernel is power-like in the dipole sizes and it allows for the splitting of a parent dipole into pair of arbitrarily large daughter dipoles. As a result the impact parameter dependence of the amplitude contains Coulomb-like power tails in impact parameter, which need to be regulated. This has to be done  by cutting off the large dipole sizes 
in the branching kernel. This is done by including a mass parameter, $m$, which accounts for the effect of color confinement. In previous works we have tested several prescriptions and found the simulation with the  simple cutoff using the theta functions gave best description of the inclusive data. We  have therefore used this scenario for the calculation in this paper.  The modified kernel which has been used for the calculation of a solution has  therefore the  form of the kernel \eqref{eq:KernLOBal} with theta functions
\begin{multline}
{\cal K} =  \bar{\alpha}_s(x_{01}^2)\left[\frac{1}{x_{02}^2}\left(\frac{\alpha_s(x_{02}^2)}{\alpha_s(x_{12}^2)} - 1\right) + \frac{1}{x_{12}^2}\left(\frac{\alpha_s(x_{12}^2)}{\alpha_s(x_{02}^2)} - 1\right) + \frac{x_{01}^2}{x^2_{12} x_{02}^2}\right] \\
\times \Theta(\frac{1}{m^2} - x_{02}^2)\Theta(\frac{1}{m^2} - x_{12}^2) \; ,
\label{eq:KernLOBalTheta}
\end{multline}
where $m$ is the mass regulator.
The value of the parameter $m$ has  been fitted to obtain the best description of the data. 
Note that the similar procedure has been utilized in Refs.~\cite{Avsar:2006jy,Flensburg:2008ag}, with the value of the parameter $r_{\rm max}=\frac{1}{m}$ to be equal around $3 \, {\rm GeV}^{-1}$.
The formula (\ref{eq:glaubermueller}) is used as an initial condition for the BK evolution for dipoles with sizes smaller than the cutoff $1/m$, for dipoles larger than the cutoff the initial condition is set to zero.

%%%%%%%%%%%%%%%%%%%%%%%%%%%%%%%%%%%%%%%%%
\subsection{Corrections to the dipole scattering amplitude and photon wave function}
\label{subsec:phenom}

There are several additional  phenomenological corrections which we have included in this calculation.  As we will see explicitly they have a non-negligible impact on the calculations. The first of these is to take into account the effect  of the skewed  gluon distribution. This is necessary as the two gluons exchanged in the production of the vector mesons need not have the same longitudinal momentum fractions $x$ and $x'$. The skewed effect vanishes at small $x$ (in the leading logarithmic limit in $\ln 1/x$), however it can be substantial correction when the energy is not very large.  As suggested in Ref.~\cite{Martin:1999wb} this correction can be taken into account by a multiplicative factor on the standard gluon distribution as follows

\be
(xg(x,\eta^2))_{sk} = xg(x,\eta^2) \frac{2^{(2\lambda_{sk} + 3)}}{\sqrt{\pi}}\frac{\Gamma(\lambda_{sk} +5/2)}{\Gamma(\lambda_{sk} +4)} \; ,
\ee
where
\be
\lambda_{sk} \equiv \frac{\partial \ln(xg(x,\eta^2))}{\partial \ln(1/x)} \; .
\ee
Strictly speaking, in the dipole formalism  there is no integrated gluon distribution, but rather the dipole scattering amplitude. However, since the 
initial condition is taken as the Glauber-Mueller model in the form (\ref{eq:glaubermueller}), with explicit dependence on  the gluon density $xg(x,\eta^2)$, we can implement
the correction for skewedness inside the  initial condition. Then the evolution to lower values of $x$ is computed
 according to Eq.~\eqref{eq:BK}.

In addition we have included the correction for the real part of the scattering amplitude.  This effect  can be taken into account by multiplying the amplitude by $(1+\beta^2)$ where
\be
\beta = \tan(\pi \lambda_{r} / 2) \; ,
\ee
with
\be
\lambda_{r} = \frac{\partial \ln(A_{T,L}^{\gamma*p \rightarrow Vp})}{\partial \ln(1/x)}.
\ee
In the above equation $\beta$ is the ratio of the real to imaginary part of the scattering amplitude.
A similar procedure was used previously in the other descriptions  of the vector meson production \cite{Nemchik:1996cw,Forshaw:2003ki,Martin:1999wb}.

Finally, a correction to the photon wave function is necessary in order to modify and enhance the contribution at low $Q^2$.   The photon light-cone wave function given in Eqs. (~\ref{eq:PhotonT}) and (\ref{eq:PhotonL})  has been derived in the perturbation theory under the assumption of the presence of hard scale $Q^2$, which in turn leads to the small dipole sizes $r$. This statement is most accurate for the longitudinal contribution, but in the case of the transverse contribution the endpoint singularities in $z$ cause the distribution in dipole sizes to be broader, even for very large scales $Q^2$.   At low values  of photon virtuality $Q^2$ the corresponding dipoles which contribute to the cross section are very large and non-perturbative. Consequently the cross section with which they interact with the proton is also large.  Therefore,  at these large dipole sizes the photon has a hadronic component, which is non-perturbative.  To account for this effect we use a modification of the photon wave function in a way suggested in Ref.~\cite{Forshaw:2003ki} and also used in Ref.~\cite{Flensburg:2008ag}

\be
|\Psi_{\gamma} |^2 \rightarrow |\Psi_{\gamma}|^2 \left(\frac{1+B e^{-\omega^2 (x_{01} - R)^2}}{1+ B e^{-\omega^2 R^2}}\right).
\label{eq:WFcorrection}
\ee

The constants $B,\omega$ and $R$ are the  parameters which have to be adjusted to fit the data. This factor provides an enhancement for dipoles which have a  hadronic size. The numerical values of the parameters  $B,\omega,R$ which  are used in the calculation are given in Table~\ref{Table:Parameters}.

%%%%%%%%%%%%%%%%%%%%%%%%%%%%%%%%%%%%%%%%%
\begin{center}
\begin{table}
\begin{center}
\begin{tabular}{| l || c |}
\hline
$\mu_0$ & 1.16547 ${\rm GeV}$\\
$D$ & 2\\
$\mu$ & 0.52 ${\rm GeV}$\\
$A_g$ & 2.55042\\
$\lambda_g$ & 0.01980\\
\hline
\end{tabular}
\hspace*{1cm}
\begin{tabular}{| l || c |}
\hline
$B_G$ & 3.65 ${\rm GeV}^{-2}$\\
$m$ & 0.37 ${\rm GeV}$\\
$R$ & 6.8 ${\rm GeV}^{-1}$\\
$B$ & 6.0\\
$\omega^2$ & 0.2 ${\rm GeV}^2$\\
\hline
\end{tabular}
\end{center}
\caption{Values of the free parameters used in the calculations.}
\label{Table:Parameters}
\end{table}
\end{center}

%%%%%%%%%%%%%%%%%%%%%%%%%%%%%%%%%%%%%%%%%
\subsection{Vector meson wave function}
\label{eq:vecmes}

Several different models
for the vector meson wave function exist in the literature,  for example see
\cite{Dosch:1996ss,Kulzinger:1998hw,Nemchik:1994fp,Nemchik:1996cw,Brodsky:1994kf,Frankfurt:1995jw,deTeramond:2005su}.  Typically, there are many  uncertainties in obtaining the wave functions for the vector mesons, however they are constrained by model independent features. First of all, $\psi_V^{h,\bar{h}}$ has to satisfy the following normalization condition 

\begin{equation}
1=\sum_{h,\bar h}{\int d^2{\mathbf r}\,dz\,|\psi^{h,\bar
h}_V(z,{\mathbf r})|^2}\ .
\label{eq:norm}
\end{equation}

In addition, the value of the wave function at the origin is related
to the leptonic decay width $\Gamma(V\rightarrow e^+ e^-)$ of the vector meson,
given by the following formula

\begin{equation}
\int_0^1 dz\,\psi_V(z,r=0)=\sqrt{\frac{\pi}{N_c}}\frac{f_V}{2\hat
e_V},\quad\mbox{where}\quad
\langle0|J^\mu_{\mbox{\footnotesize em}}(0)|V\rangle\equiv
e_q f_V\,m_V\,\varepsilon^\mu\ .
\label{eq:lwid}
\end{equation}

Here $f_V$ is the coupling of the meson to the electromagnetic current and 
$\hat e_V$ the isospin factor, which is the
effective charge of the quarks in units of the elementary charge $e$:
for the $\rho$ meson it is the charge of the combination 
$(u\bar u\!-\!d\bar d)/\sqrt{2}$, i.e. $\hat{e}_V = 1/\sqrt{2}$.
Finally, one requires that the mean radius 
be consistent with the electromagnetic radius of the vector meson.

In this paper we utilize the model proposed in
 \cite{Nemchik:1994fp,Nemchik:1996cw}.
In this approach   the information from spectroscopic
models is used to constrain the long distance physics. 
One assumes that the meson is composed of a constituent
quark and antiquark which move in a harmonic oscillator potential.
This results in a wave function
which has a gaussian dependence on the spatial separation between the quarks.
Additionally, this model is supplemented by the short-distance physics
driven by QCD exchange of  hard gluons between the valence quarks
of the vector meson. Finally, a relativization technique has to be
applied to the wave function.

 The wave function of the vector meson
is given by 
\begin{equation}
\psi_V^{h,\bar{h}}(z,{\mathbf r})
 \; = \;  \delta_{h,-\bar{h}} \sqrt{N_c \over 4\pi }
{1 \over m_V z(1-z)} [m_V^2 z (1-z) - {\mathbf \nabla_r}^2 + m_q^2] \phi(r,z) \, .
\label{eq:nikowf}
\end{equation}

In this model, the radial wave function $\phi(r,z)$ 
appearing in eq.(\ref{eq:nikowf}) satisfies the following normalization 
condition:
\begin{equation}
1=\frac{N_c}{2\pi}\int_0^1\frac{dz}{z^2(1-z)^2}\int d^2{\mathbf
r}\bigg\{
m_q^2\phi^2(z,{\mathrm r})+(z^2+(1-z)^2)(\partial_r\phi(z,{\mathrm
r}))^2\bigg\}\ .
\label{eq:norm2}
\end{equation} 
This equation is essentially the normalization condition~(\ref{eq:norm})
applied to the transversely polarized meson wave function.
The function $\phi(r,z)_{(L/T)}$ is defined by:
\begin{multline}
\label{eq:psilt}
\phi(r,z)_{(L/T)}=\Psi_{0{L/T}}(1S)\\
\times\bigg\{4z(1-z)\sqrt{2\pi R^2}
\exp\left(-\frac{m_q^2 R^2}{8z(1\!-\!z)}\right)
\exp\left(-\frac{2z(1\!-\!z)r^2}{R^2}\right)\exp\left(\frac{m_q^2 R^2}{2}\right)
\bigg\}\ .
\end{multline}
The masses of the light quarks are taken to be $0.14\;\mbox{GeV}$ and the charm mass is taken to be $1.4\;\mbox{GeV}$.

The other two parameters $\Psi_{0{L/T}}(1S)$ and $R^2$ are chosen by
taking into account several constraints: the normalization condition (\ref{eq:norm2})
for the wave function has to be satisfied, and the value of the
leptonic decay width must agree with the experimental measurement.
Additionally, the mean radius of the meson has to be of the order of a hadronic scale.  The normalization $\Psi_{0{L/T}}(1S)$ is different  for the longitudinal and transverse wave functions of each vector meson. On the other hand   the value of $R^2$ is the same for each polarization but different for each vector meson.  The values of the parameters used for the vector meson wave functions can be found in Table~\ref{Table:VMparam}.

%%%%%%%%%%%%%%%%%%%%%%%%%%%%%%%%%%%%%%%%%
\begin{center}
\begin{table}
\begin{center}
\begin{tabular}{| l || c | c | c |}
\hline
 & $\rho$ &  $\phi$ & $J/\Psi$\\
\hline
$\Psi_0(1S)_T$ & 0.025 & 0.028 & 0.039\\
$\Psi_0(1S)_L$ & 0.024 & 0.025 & 0.039\\
$R^2$ & 12.77 & 11.0  & 2.188\\
\hline
\end{tabular}
\end{center}
\caption{Values of the free parameters used in the vector meson wave functions.}
\label{Table:VMparam}
\end{table}
\end{center}
%%%%%%%%%%%%%%%%%%%%%%%%%%%%%%%%%%%%%%%%%

%%%%%%%%%%%%%%%%%%%%%%%%%%%%%%%%%%%%%%%%%
\section{Comparison with the experimental data}
\label{sec:results}

In this section we compare the results of our calculation based on the  numerical solution of the BK equation with the impact parameter dependence to the  data from H1 and ZEUS on $\rho,\phi,$ and $J/\Psi$ production.

 As discussed above, the initial condition for the evolution in $x$ is given by Eq.~(\ref{eq:glaubermueller}). This formula   contains several free parameters, such as  the value of  $B_G$, the parameters $A_g,\lambda_g$ in the initial gluon distribution $xg(x,\eta^2) = A_g x^{-\lambda_g} (1-x)^{5.6}$,  and the parameters $\mu_0$ and $D$ which enter into the definition of the scale  $\eta^2 = \frac{D^2}{r^2} + \mu_0^2$.  The initial impact parameter profile $T(b)$ (\ref{eq:profile}) is assumed to be gaussian.  This profile is substantially changed during the course of the evolution with $x$. In addition, the kernel contains two scales:  the scale $\mu$ in the running of the coupling as well as the mass parameter $m$.  All these parameters enter into the calculation of the dipole scattering amplitude which is found by the iterative solution of the Eq.~\eqref{eq:BK}.  The    computational time that is required   to obtain the solution for each set of parameter values  is rather long (on the order of 24 hrs on 32 cores). Therefore we have chosen to vary only a subset of these parameters.   Parameters $A,\lambda_g,C,\mu_0$  were fixed and  their values were taken from Ref.~\cite{Kowalski:2006hc}.  The only parameter which was varied in the initial condition Eq.~(\ref{eq:glaubermueller}) was $B_G$.  The value of $B_G$  was adjusted  to obtain the best description of the $t$ dependence of the differential cross section $d\sigma/dt$.
 In particular this parameter controls  the magnitude of the experimentally determined value $B_D$, which is the slope of the $t$ dependence, i.e. $d\sigma/dt\sim \exp(-B_D|t|)$. In addition,    the mass parameter in the kernel $m$,  which controls the $W$ dependence of the slope $B_D$  was also fitted, together with $B_G$ in Eq.~\eqref{eq:glaubermueller}, to obtain the best description of the data.  In our analysis we set $m = \frac{1}{\sqrt{2 B_G}}$ but these two parameters need not be directly linked.
 
There are also  parameters that are present in the formulae for the cross section for the production of the various vector mesons.  These include the quark masses as well as the parameters in the vector meson wave function and  non-perturbative correction to the photon wave function. The variation of these parameters does not cost so much computational  time therefore a larger number of them were varied for the best fit.  The complete set of parameters used in this paper can be found in Table \ref{Table:Parameters}.

Using the framework described above we have performed a fit of the calculation to the inclusive data on the structure functions in DIS. Note that this was necessary as  the  details of the procedure presented above are slightly different than that originally used in \cite{Berger:2011ew}.  Good agreement between the inclusive data and the calculation was found using this procedure. 

%%%%%%%%%%%%%%%%%%%%%%%%%%%%%%%%%%%%%%%%%
\begin{figure}
\centering
\includegraphics[angle=270,width=0.48\textwidth]{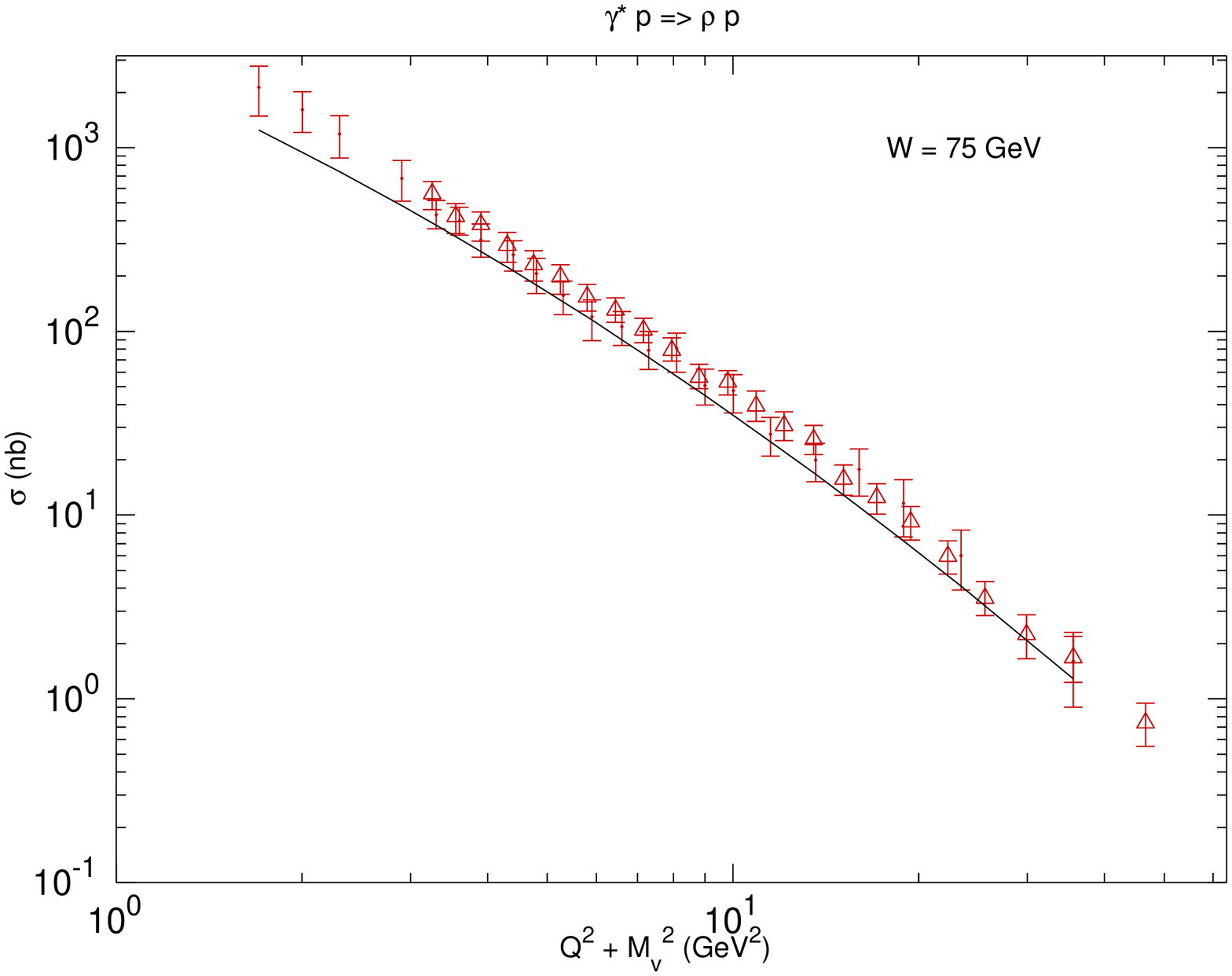}
\includegraphics[angle=270,width=0.48\textwidth]{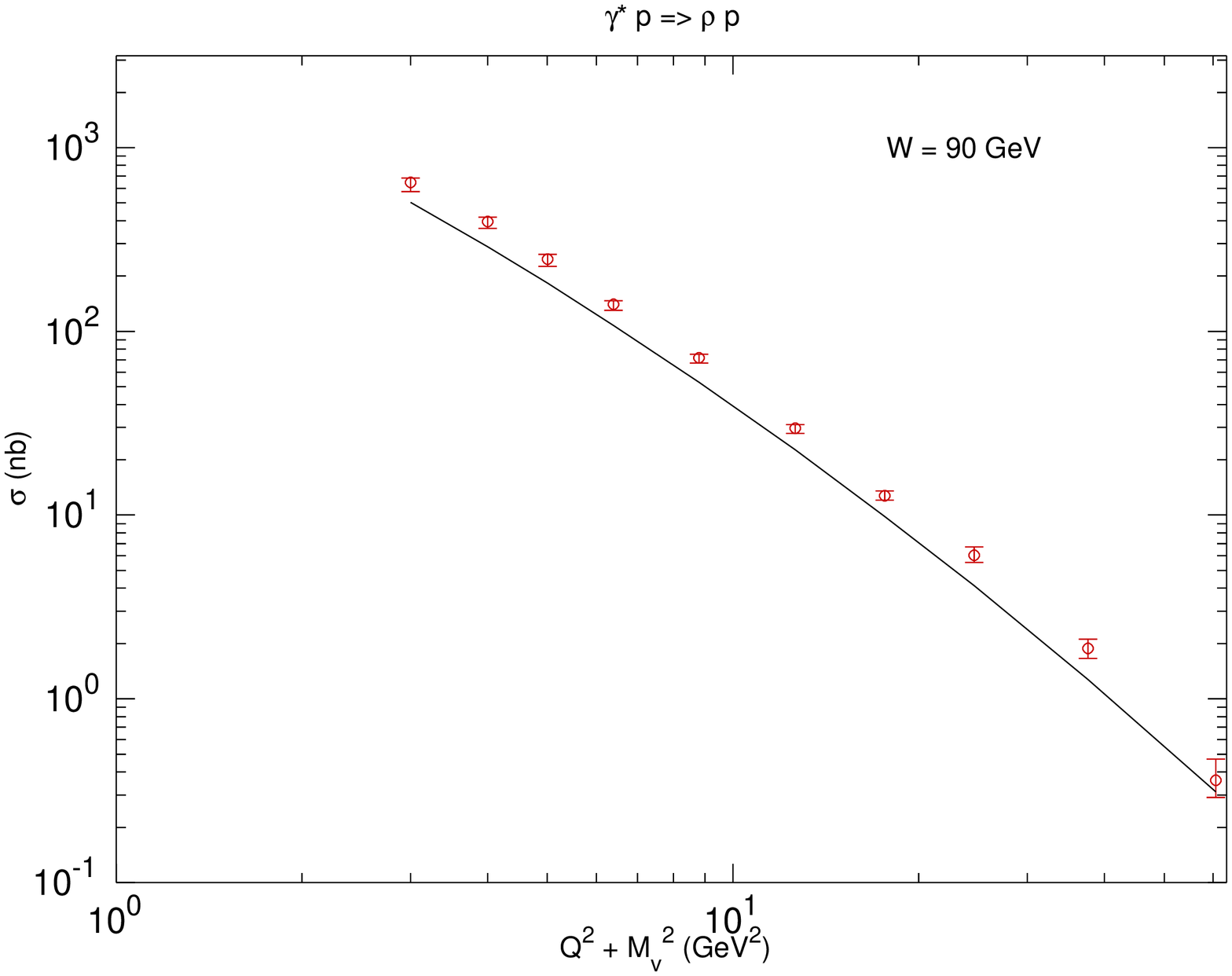}
\caption{Cross section $\sigma(Q^2,W^2)$ for the vector meson production plotted as a function of $(Q^2 + M^2_V)$ for $\rho$  elastic production. The experimental data are from \cite{Adloff:1999kg,Chekanov:2007zr,Aaron:2010}. }
\label{fig:Q2M2a}
\end{figure}
%%%%%%%%%%%%%%%%%%%%%%%%%%%%%%%%%%%%%%%%%
\begin{figure}
\centering
\includegraphics[angle=270,width=0.46\textwidth]{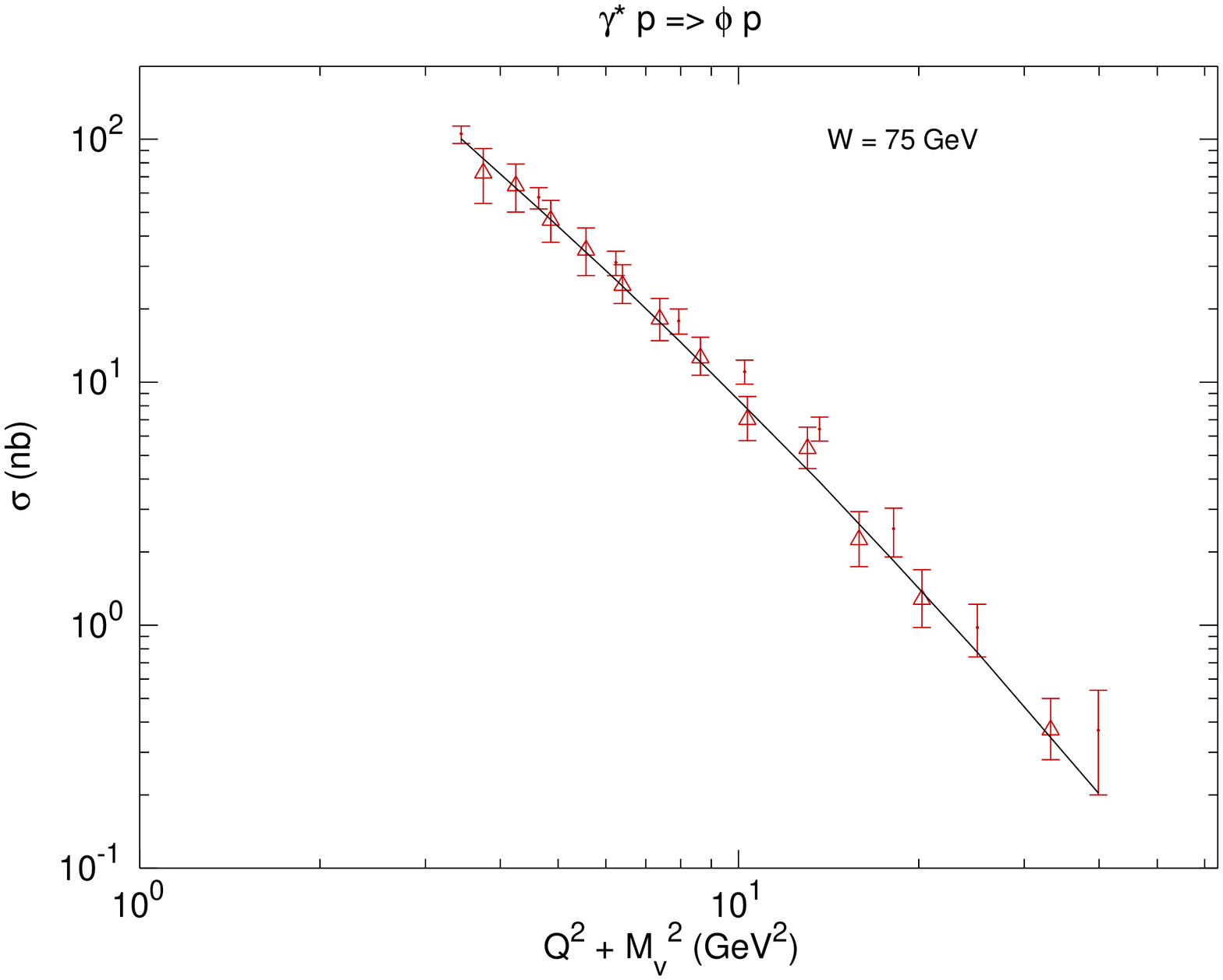}
\includegraphics[angle=270,width=0.49\textwidth]{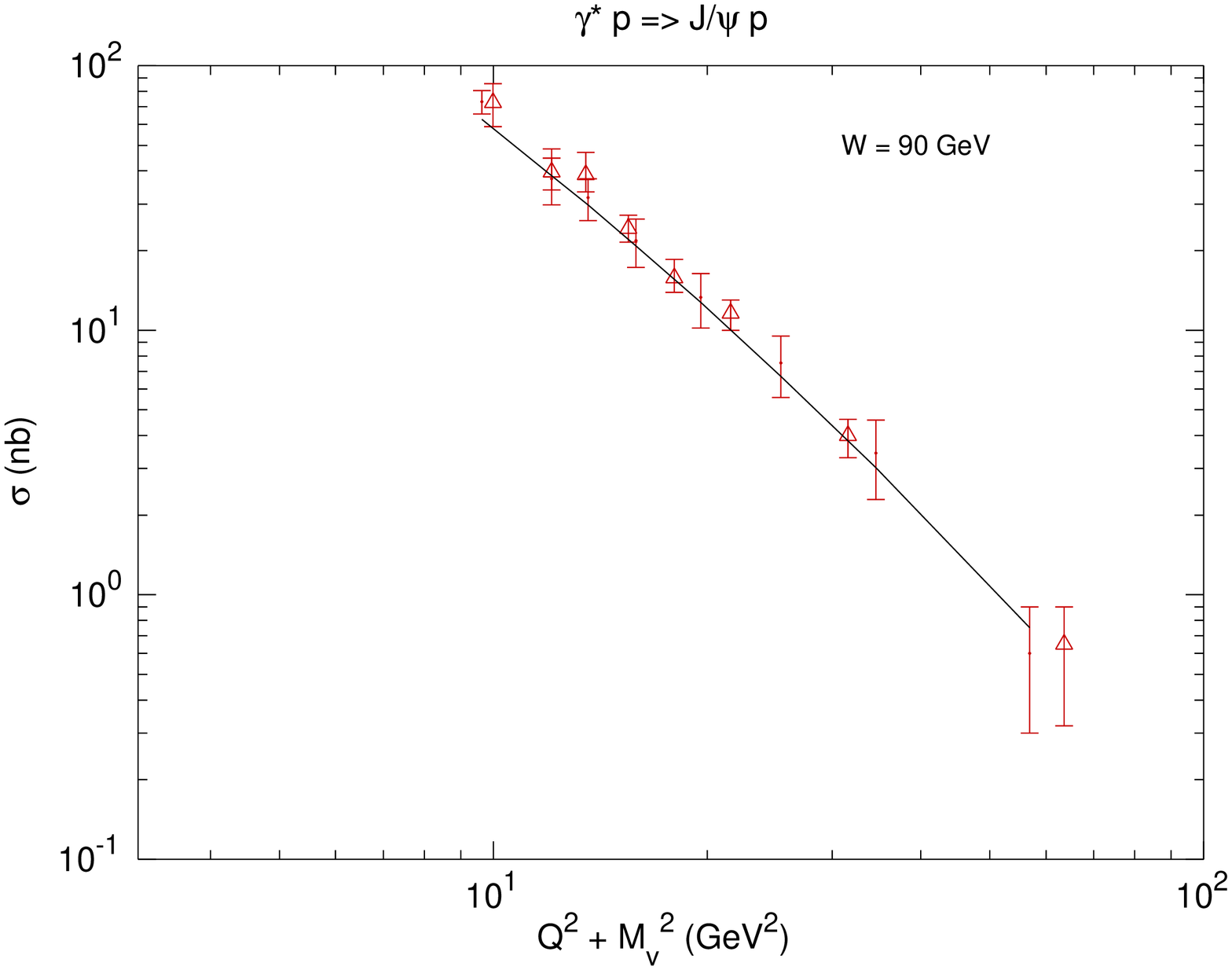}
\caption{Cross section $\sigma(Q^2,W^2)$ for the vector meson production plotted as a function of $(Q^2 + M^2_V)$ for  $\phi$ (left plot) and $J/\psi$ (right plot) elastic production. The experimental data are from \cite{Chekanov:2004mw,Chekanov:2005cqa,Aktas:2005xu,Aaron:2010}. }
\label{fig:Q2M2b}
\end{figure}
%%%%%%%%%%%%%%%%%%%%%%%%%%%%%%%%%%%%%%%%%

%%%%%%%%%%%%%%%%%%%%%%%%%%%%%%%%%%%%%%%%%
\subsection{Cross sections for exclusive vector meson production integrated over $t$}

Let us  first show the comparison between the calculation based on the dipole model with BK equation and the experimental data on the cross section for the  process of exclusive diffractive electroproduction of vector mesons, where the cross section has been integrated over the momentum transfer $t$.
The experimental data     from H1 and ZEUS for $\rho$ \cite{Adloff:1999kg,Chekanov:2007zr,Aaron:2010},  $\phi$  \cite{Chekanov:2005cqa,Aaron:2010} and  $J/\Psi$ \cite{Chekanov:2004mw,Aktas:2005xu} were used.

Figures \ref{fig:Q2M2a} and \ref{fig:Q2M2b} shows the  cross section for production of  $\rho$, $\phi$  and $J/\Psi$ vector mesons as a function of variable $(Q^2+M_V^2)$, where $M_V^2$ is the mass squared of the corresponding vector meson.  This variable is commonly used instead of $Q^2$ itself as it provides the scale for the vector meson. One cannot, however, expect that cross sections for different vector mesons will behave identically when this variable is used. We therefore show the cross sections for different species of vector mesons separately. The momentum transfer $t$ has been integrated in the ranges provided by the experimental data. For the data on $\rho$ production from \cite{Adloff:1999kg} the range in $t$ is $|t|<0.5 \; {\rm GeV}^2$ and  $|t|<1.0 \; {\rm GeV}^2$ for data from \cite{Chekanov:2007zr}; $|t|<3 \; {\rm GeV}^2$ for data from \cite{Aaron:2010}. For the data on $\phi$ production from \cite{Chekanov:2005cqa} the range is $|t|<0.6 \; {\rm GeV}^2$ and $|t|<3 \; {\rm GeV}^2$ for data from \cite{Aaron:2010}. Finally for $J/\Psi$ the range of $|t|<1.2 \; {\rm GeV}^2$ in \cite{Aktas:2005xu} and $|t|<1.0 \; {\rm GeV}^2$ for data from \cite{Chekanov:2004mw}. 
The theoretical curves shown in Figs.~\ref{fig:Q2M2a},\ref{fig:Q2M2b} are evaluated  for constant energy $W$ where $W=90 \; {\rm GeV}$ for the $J/\Psi$ cross section and $W=75 \; {\rm GeV}$ or $W=90\; {\rm GeV}$ for the $\phi$ and $\rho$ cross sections depending on the  data set used.

%%%%%%%%%%%%%%%%%%%%%%%%%%%%%%%%%%%%%%%%%
\begin{figure}
\begin{center}
\includegraphics[angle=270,width=0.49\textwidth]{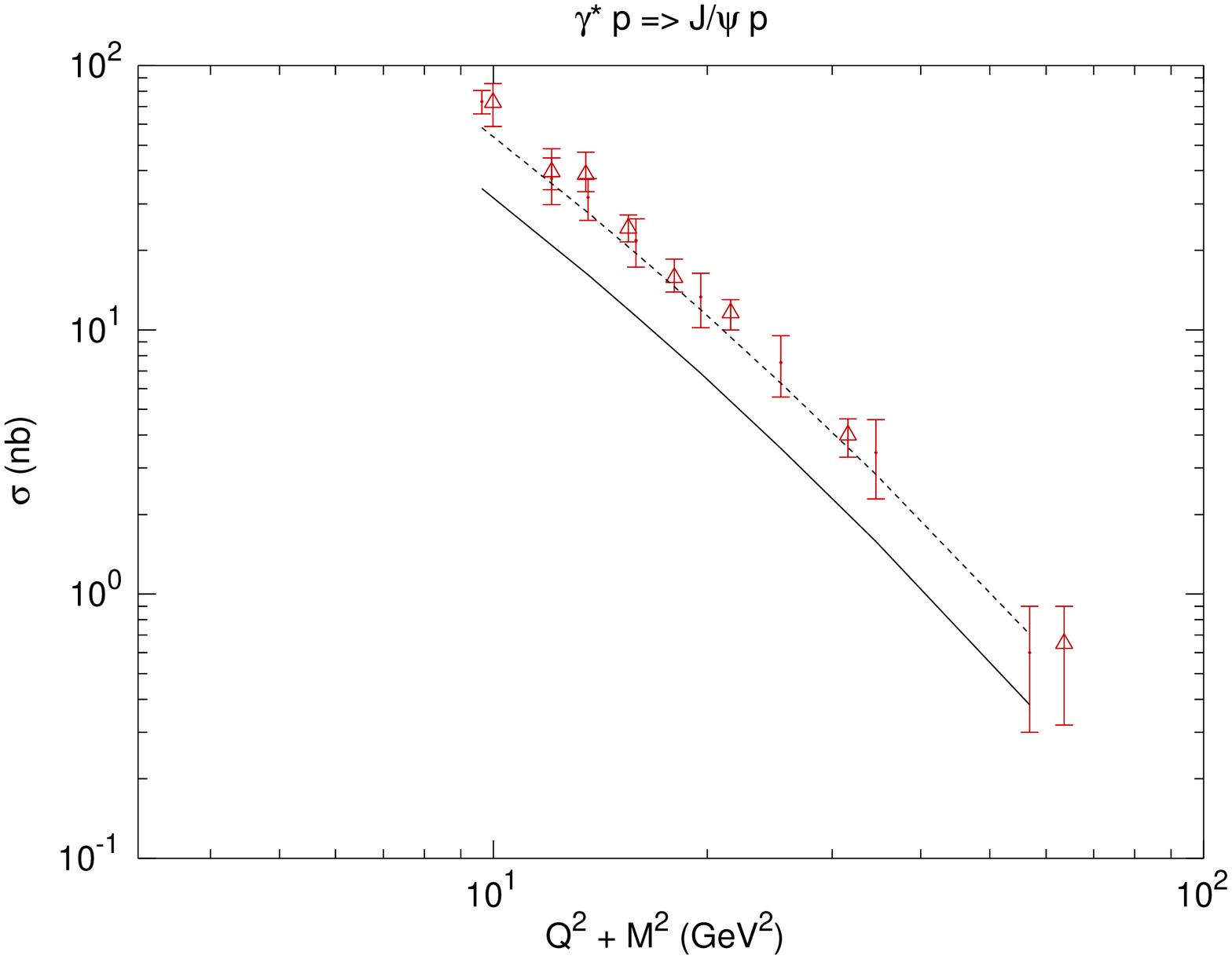}
\includegraphics[angle=270,width=0.45\textwidth]{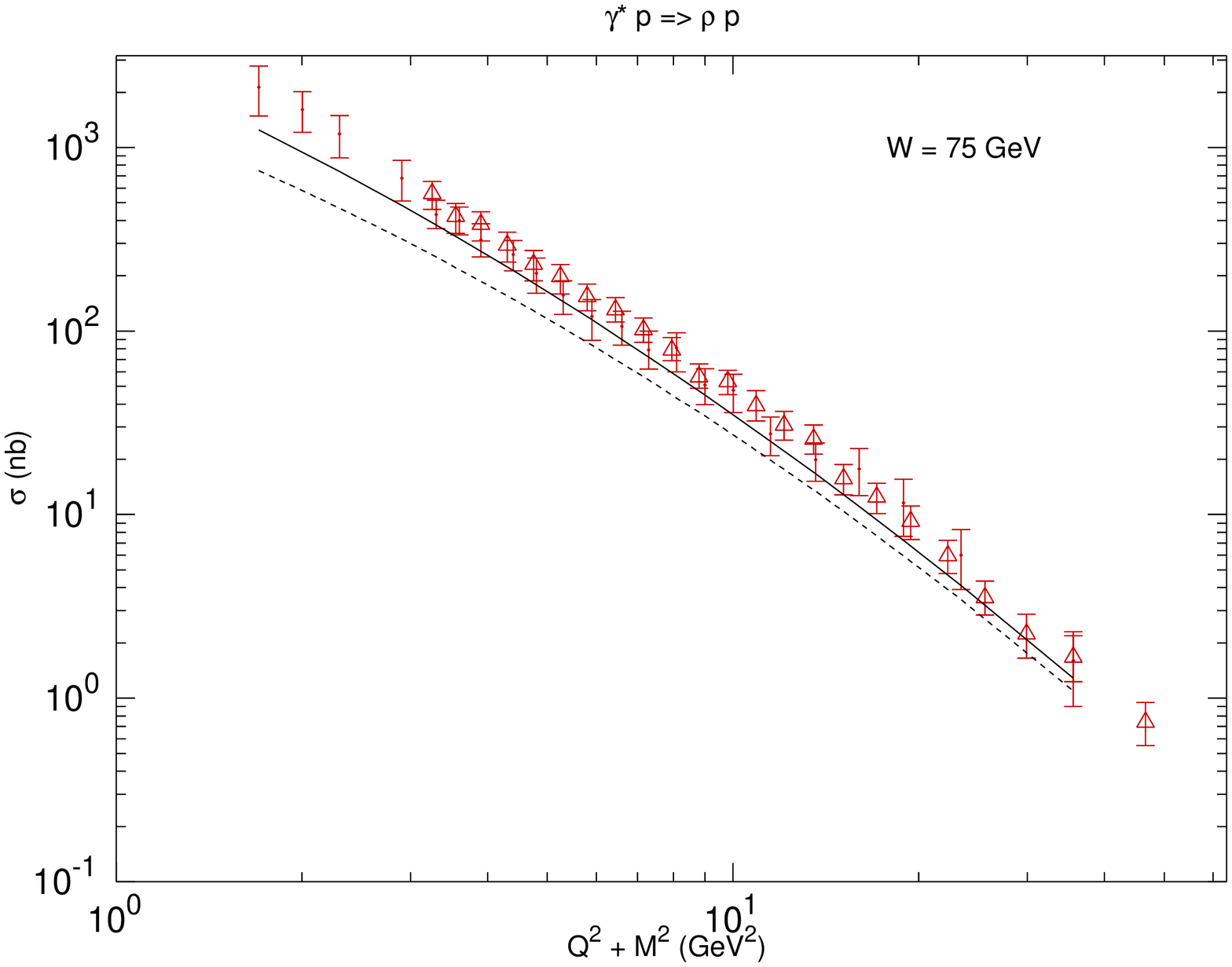}
\caption{Illustration of the effect of various corrections on the integrated cross section for the vector meson production. Left: the solid line represents the calculation without skewedness and the real part of the scattering amplitude, while the dashed line represents calculation which includes skewedness and a correction to account for the real part of the scattering amplitude in the initial condition.  No correction to the photon wave function was made on either of these curves. Right: both curves have the initial condition corrected for the real part of the scattering amplitude and include a skewed gluon distribution correction.  The solid line represents the inclusion of the photon wave function correction and the dashed line is the calculation without the photon wave function correction term. }
\label{fig:corrections}
\end{center}
\end{figure}
%%%%%%%%%%%%%%%%%%%%%%%%%%%%%%%%%%%%%%%%%

We have tested the sensitivity of the cross section due to  the inclusion of the various corrections discussed in Sec.~\ref{subsec:phenom}. This is shown in two plots in  Fig.~\ref{fig:corrections}.
In the left plot  the effect of including the correction due to the skewed effect in the initial condition is shown with data for $J/\Psi$ production. We see that the effect is the substantial change in the normalization
of the cross section, without distortion of the $Q^2$ dependence, and calculation without skewed effect correction completely fails to match the data. We note that even with the substantial variation of the free parameters it would be difficult to match the normalization of the cross section without this additional effect. 

In the plot on the right hand side we  demonstrate the effect of including the non-perturbative modification of the
photon wave function in the form of Eq.~\eqref{eq:WFcorrection}.  This correction is especially important for low values of $Q^2$ and light vector mesons. This is to be expected as this correction vanishes for large $Q^2$ in order to reproduce the perturbative expression for the photon wave function. Thus this correction has a negligible effect
at large values of $Q^2>20 \;{\rm GeV}^2$ and the cross section for $J/\Psi$ production.

\begin{figure}
\centering
\includegraphics[angle=270,width=0.45\textwidth]{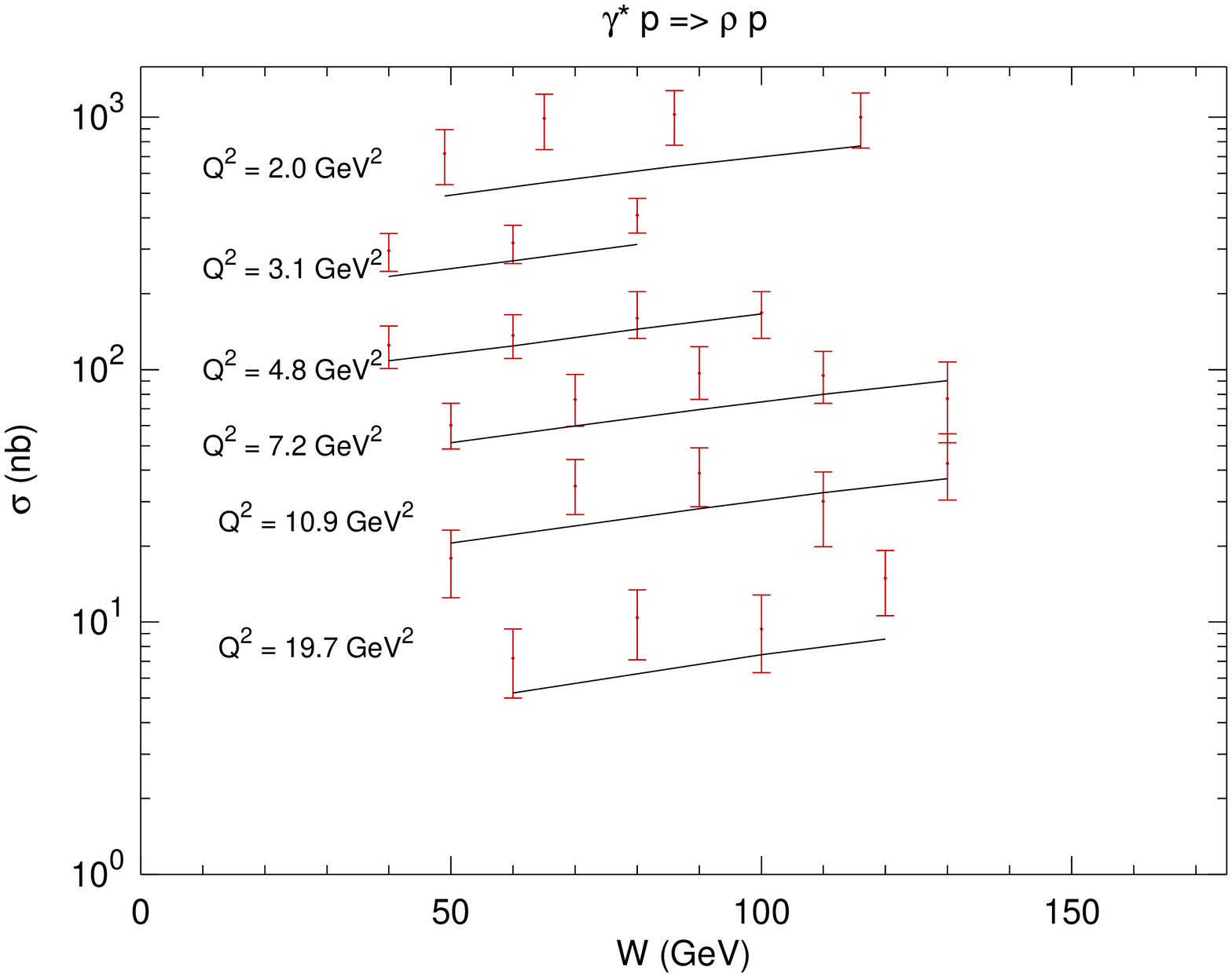}
\includegraphics[angle=270,width=0.45\textwidth]{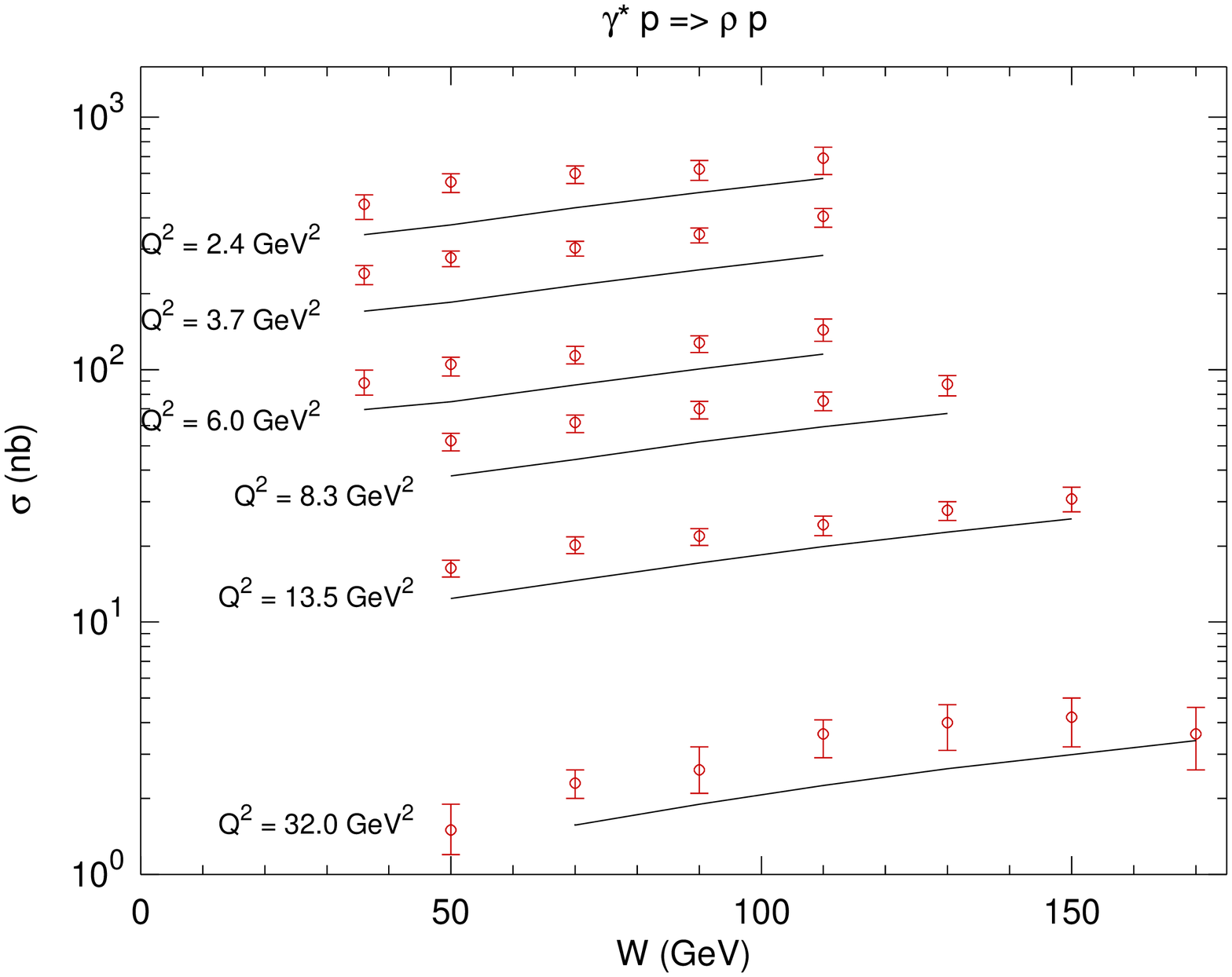}
\includegraphics[angle=270,width=0.45\textwidth]{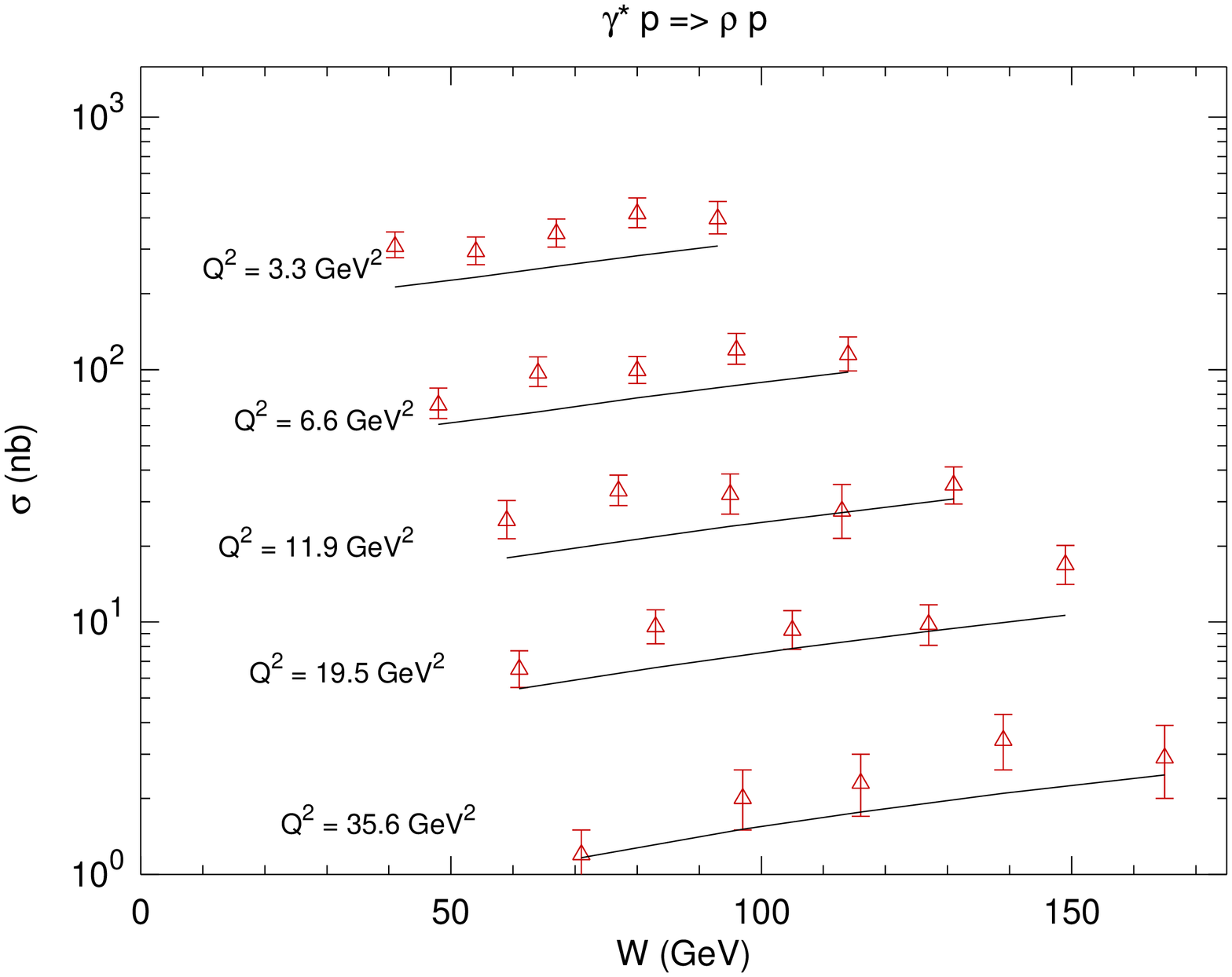}
\caption{$W$ dependence of the   vector meson cross section for elastic production of $\rho$. The experimental data are from \cite{Adloff:1999kg,Chekanov:2007zr,Chekanov:2005cqa,Aaron:2010}.  }
\label{fig:Wa}
\end{figure}

\begin{figure}
\centering
\includegraphics[angle=270,width=0.45\textwidth]{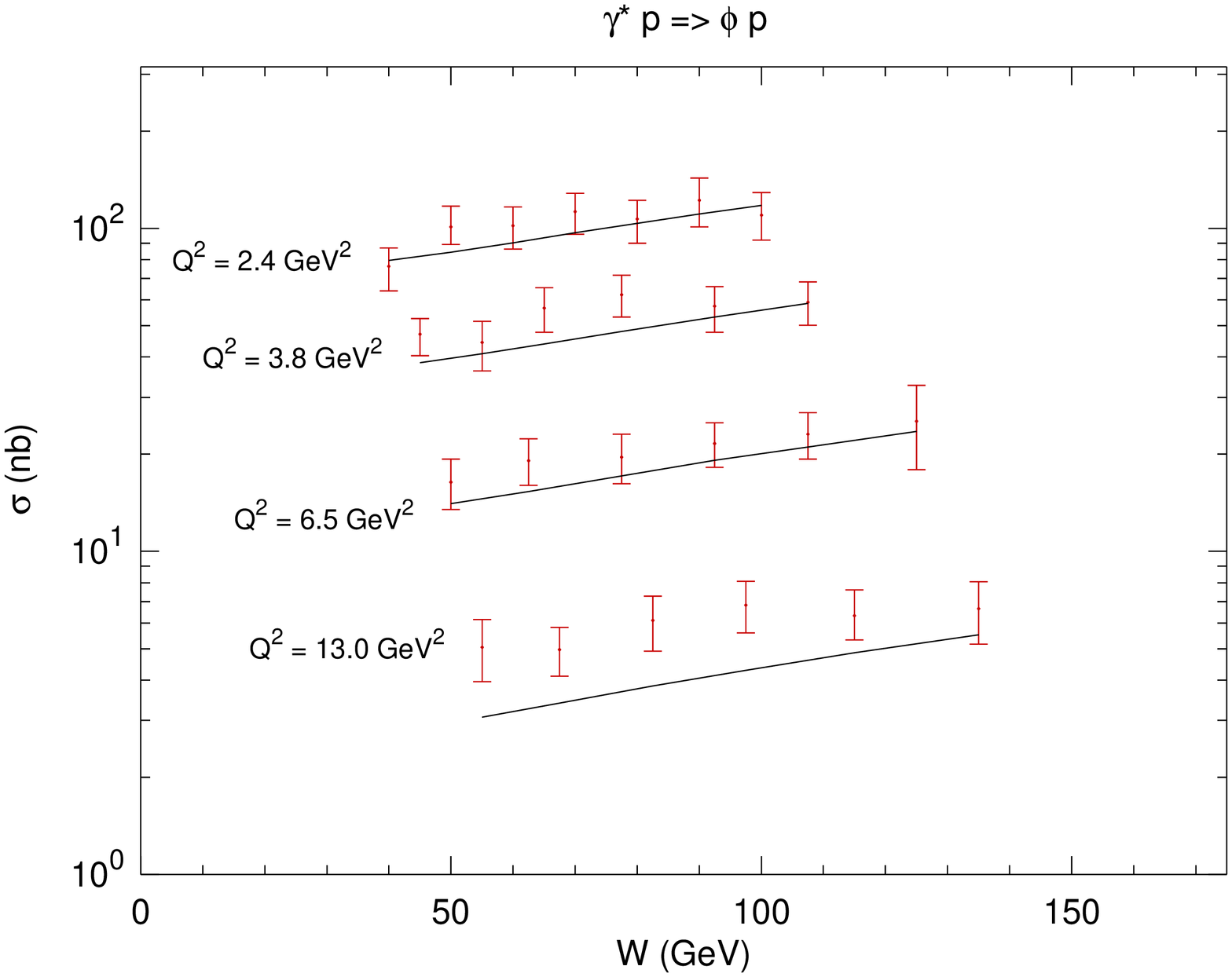}
\includegraphics[angle=270,width=0.45\textwidth]{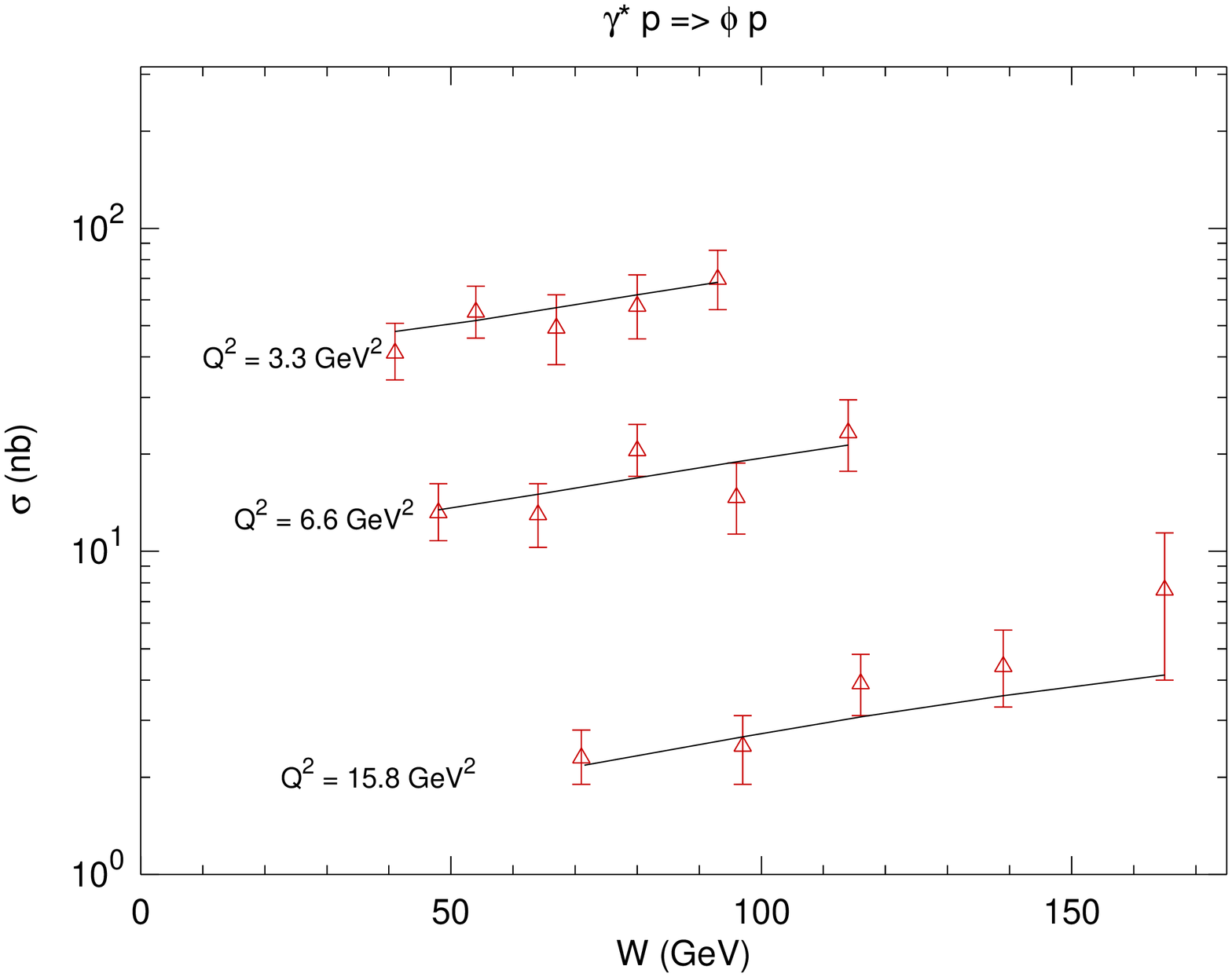}
\caption{$W$ dependence of the   vector meson cross section for elastic production of $\phi$.
The experimental data are from \cite{Chekanov:2005cqa,Aaron:2010}.
}
\label{fig:Wb}
\end{figure}

\begin{figure}
\centering
\includegraphics[angle=270,width=0.45\textwidth]{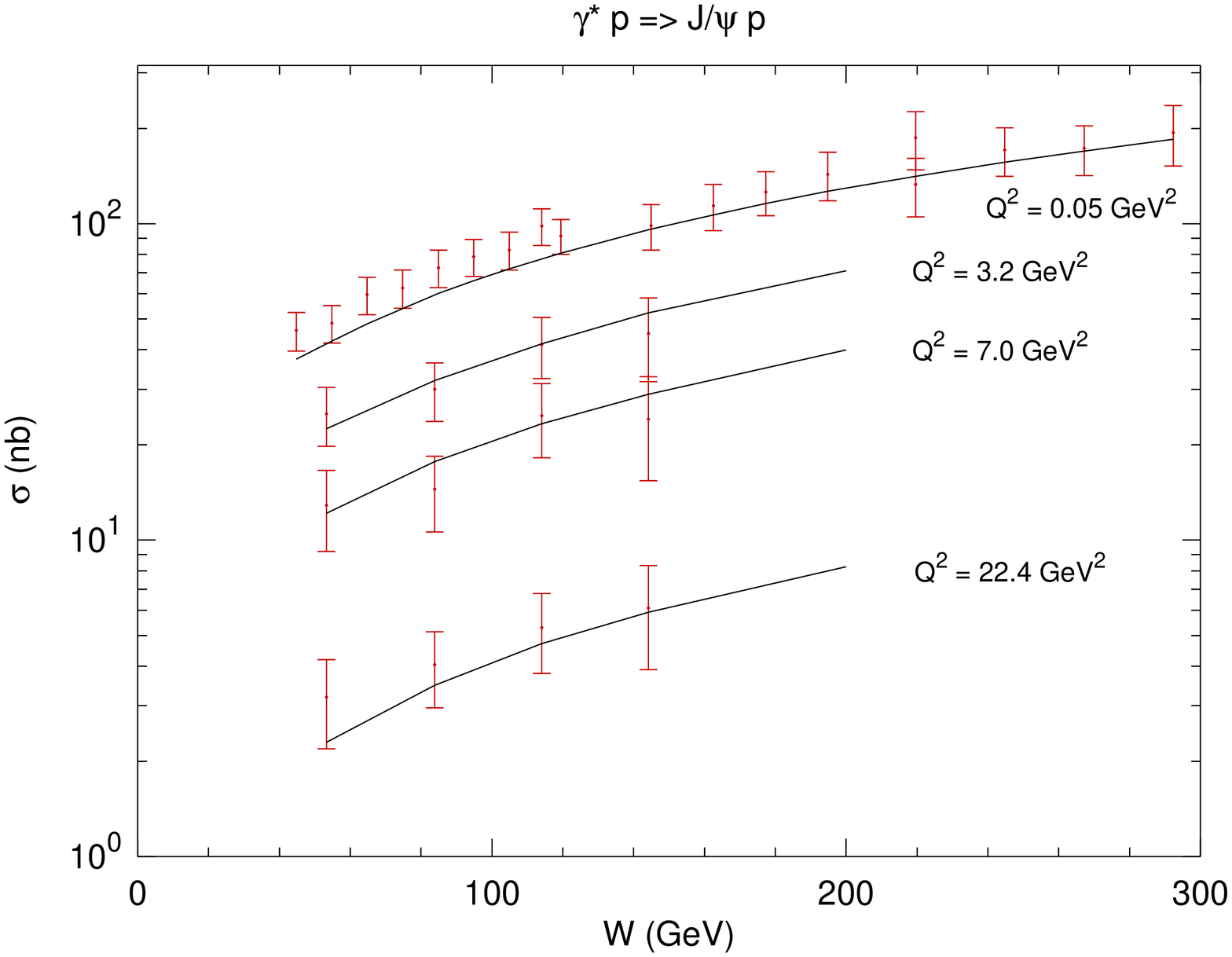}
\includegraphics[angle=270,width=0.45\textwidth]{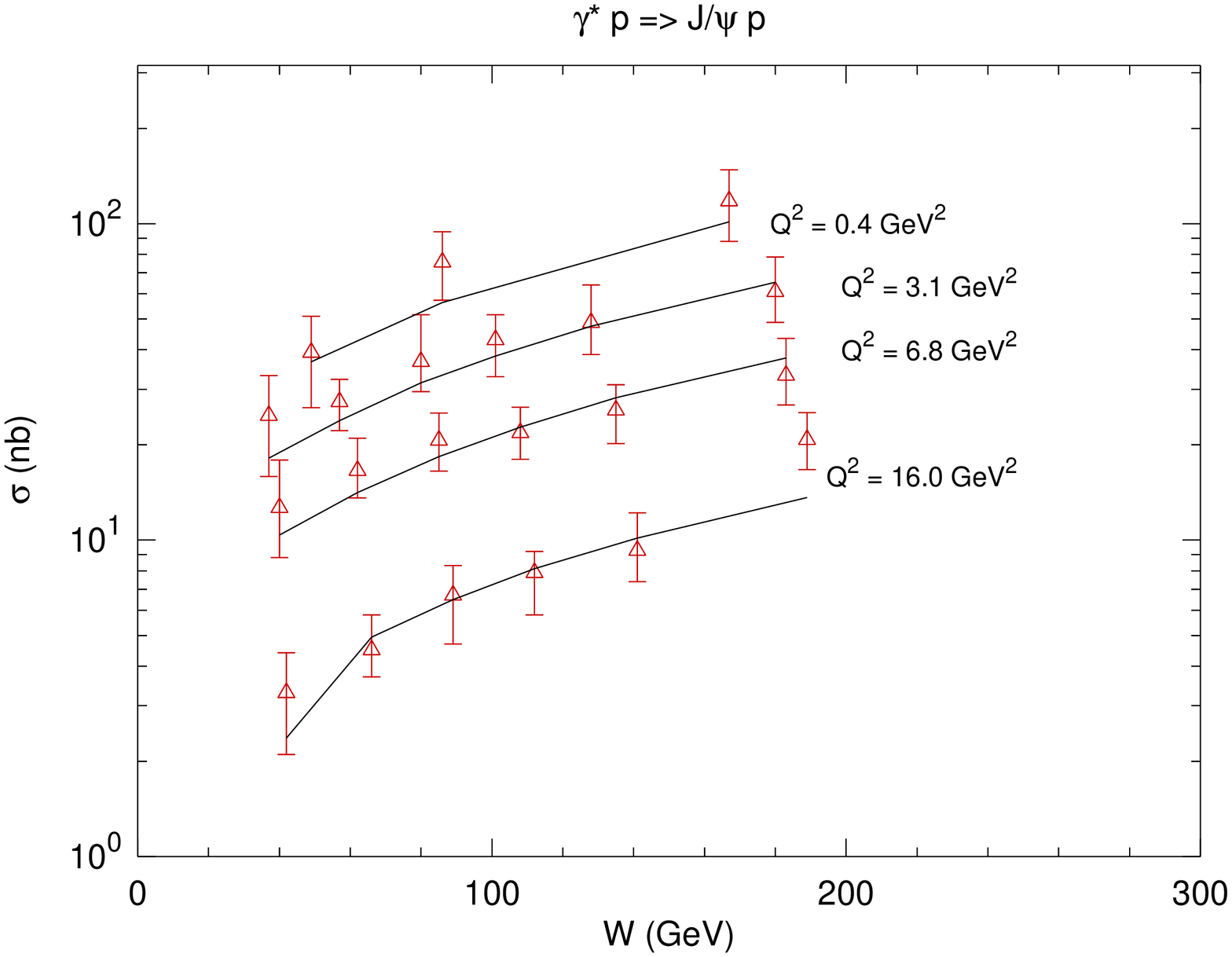}
\caption{$W$ dependence of the   vector meson cross section for elastic production of $J/\Psi$.
The experimental data are from \cite{Chekanov:2004mw,Aktas:2005xu}.}
\label{fig:Wc}
\end{figure}

The $W$ energy dependence of the  cross sections is shown in Figs.~\ref{fig:Wa},\ref{fig:Wb},\ref{fig:Wc} for $\rho,\phi,J/\Psi$ respectively. The different curves are plotted in different bins of $Q^2$.     Overall trend in both cases, i.e. for $Q^2$ and $W$ dependence, is such that the calculations describe the dependence in $Q^2$ and $W$ dependence very well for the case of $\phi$ and $J/\Psi$. The data for $\rho$ production are not well described, in particular   the normalization in this case is systematically little bit low, especially for lowest values of $Q^2$. This region is however the one which is not under perturbative control and some unknown non-perturbative corrections, related for example to the exact form of the wave function,  may play an important role in this region.

%%%%%%%%%%%%%%%%%%%%%%%%%%%%%%%%%%%%%%%%%
%%%%%%%%%%%%%%%%%%%%%%%%%%%%%%%%%%%%%%%%%
%%%%%%%%%%%%%%%%%%%%%%%%%%%%%%%%%%%%%%%%%
%%%%%%%%%%%%%%%%%%%%%%%%%%%%%%%%%%%%%%%%%

\begin{figure}
\centering
\includegraphics[angle=270,width=0.45\textwidth]{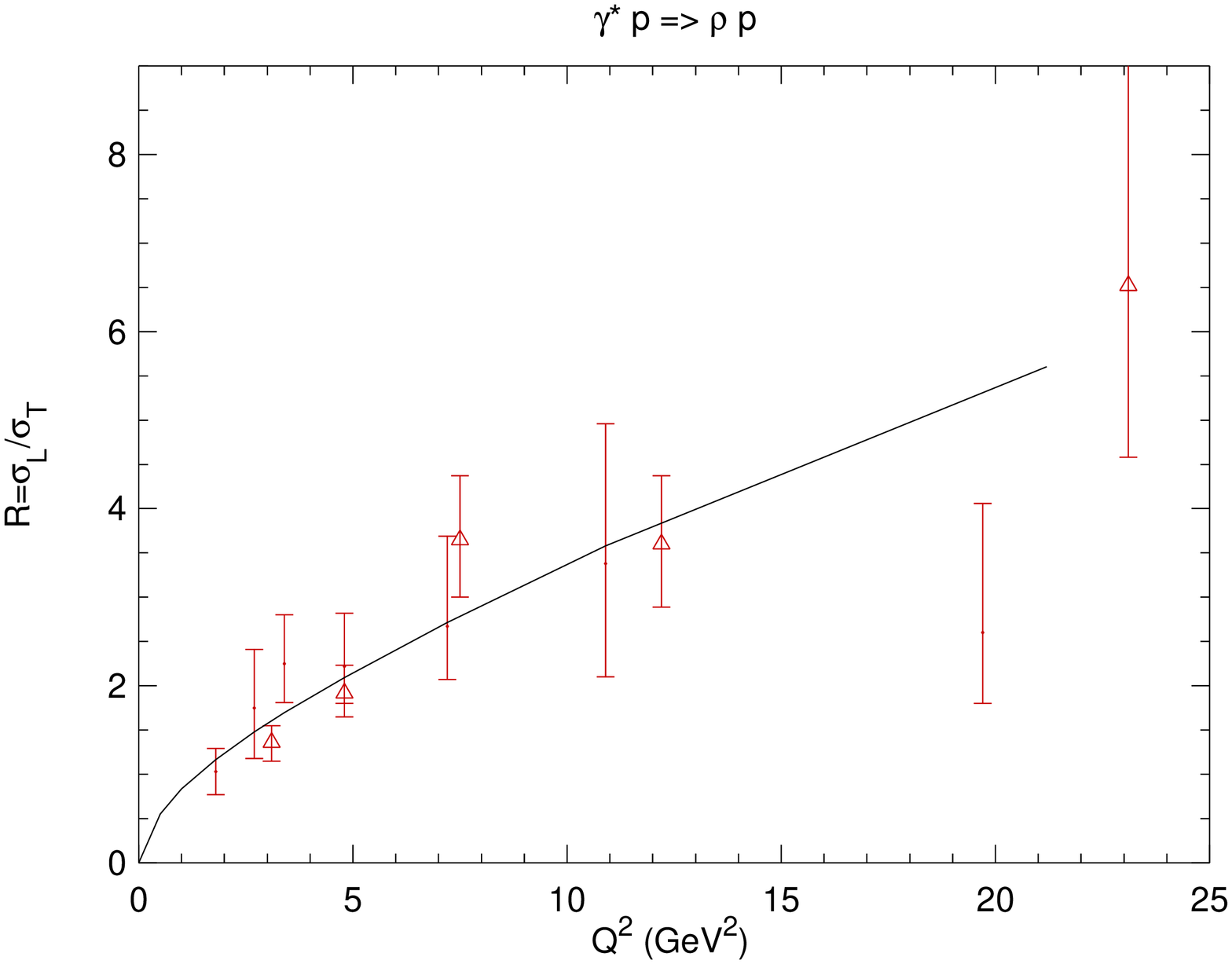}
\includegraphics[angle=270,width=0.45\textwidth]{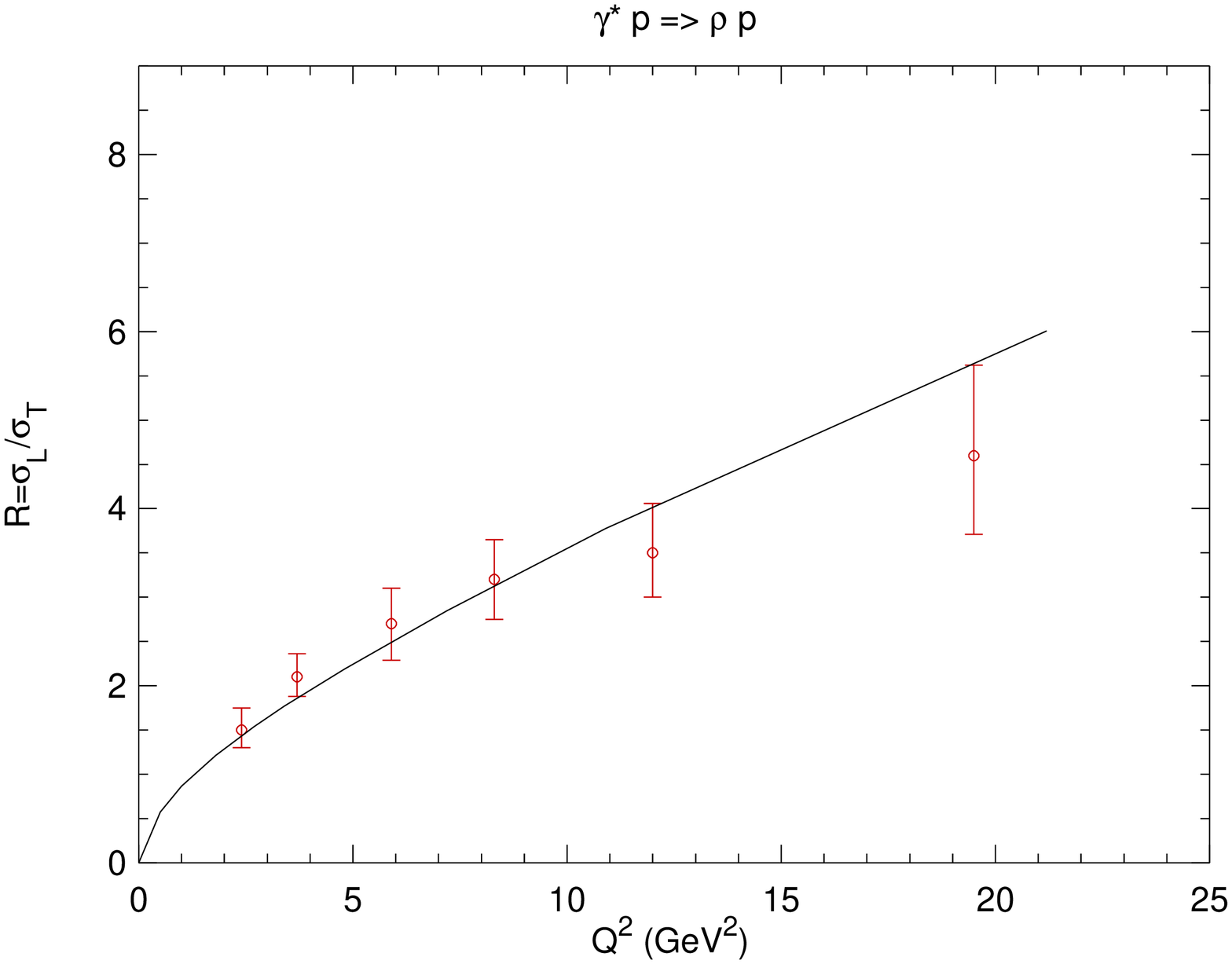}
\includegraphics[angle=270,width=0.45\textwidth]{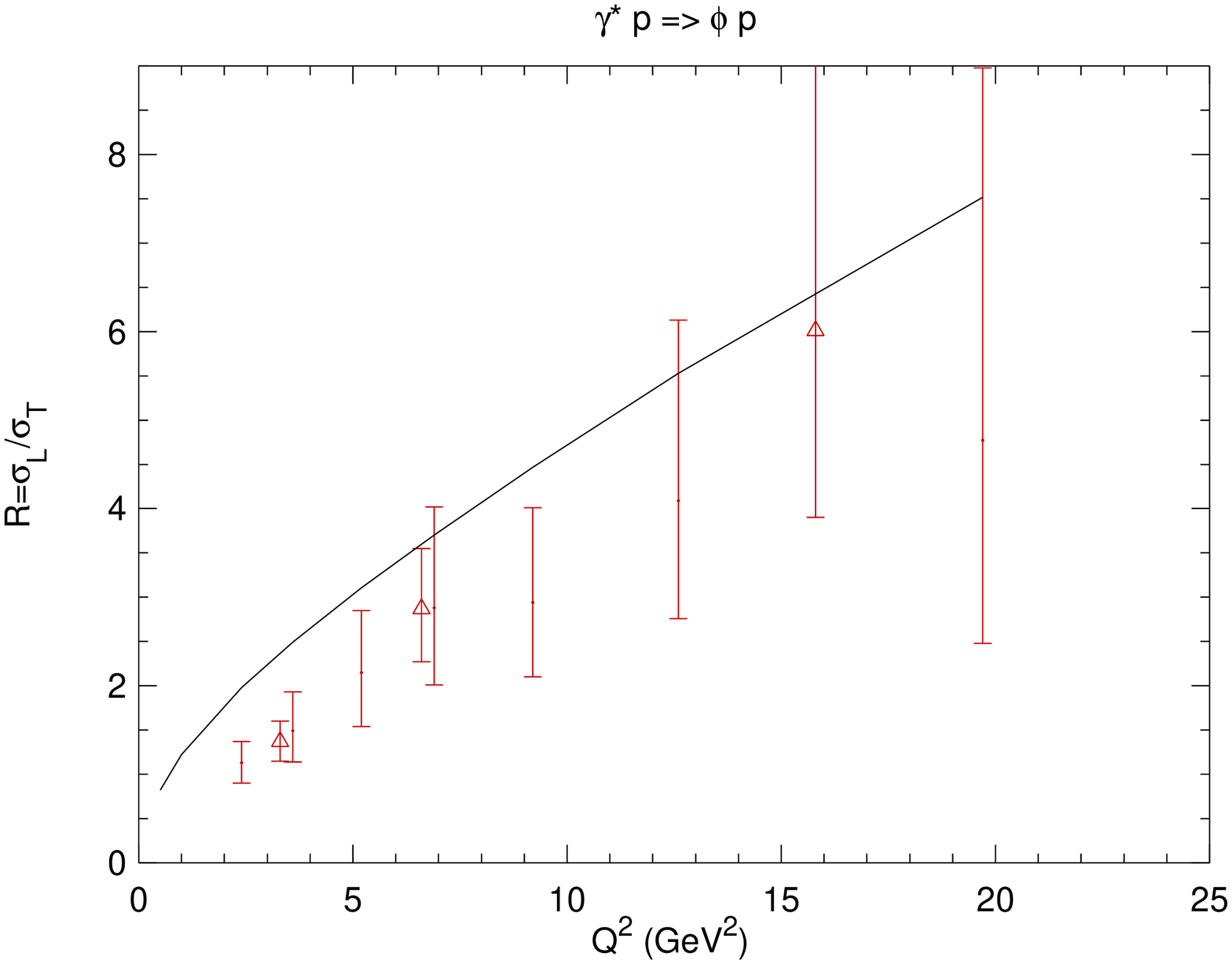}
\includegraphics[angle=270,width=0.45\textwidth]{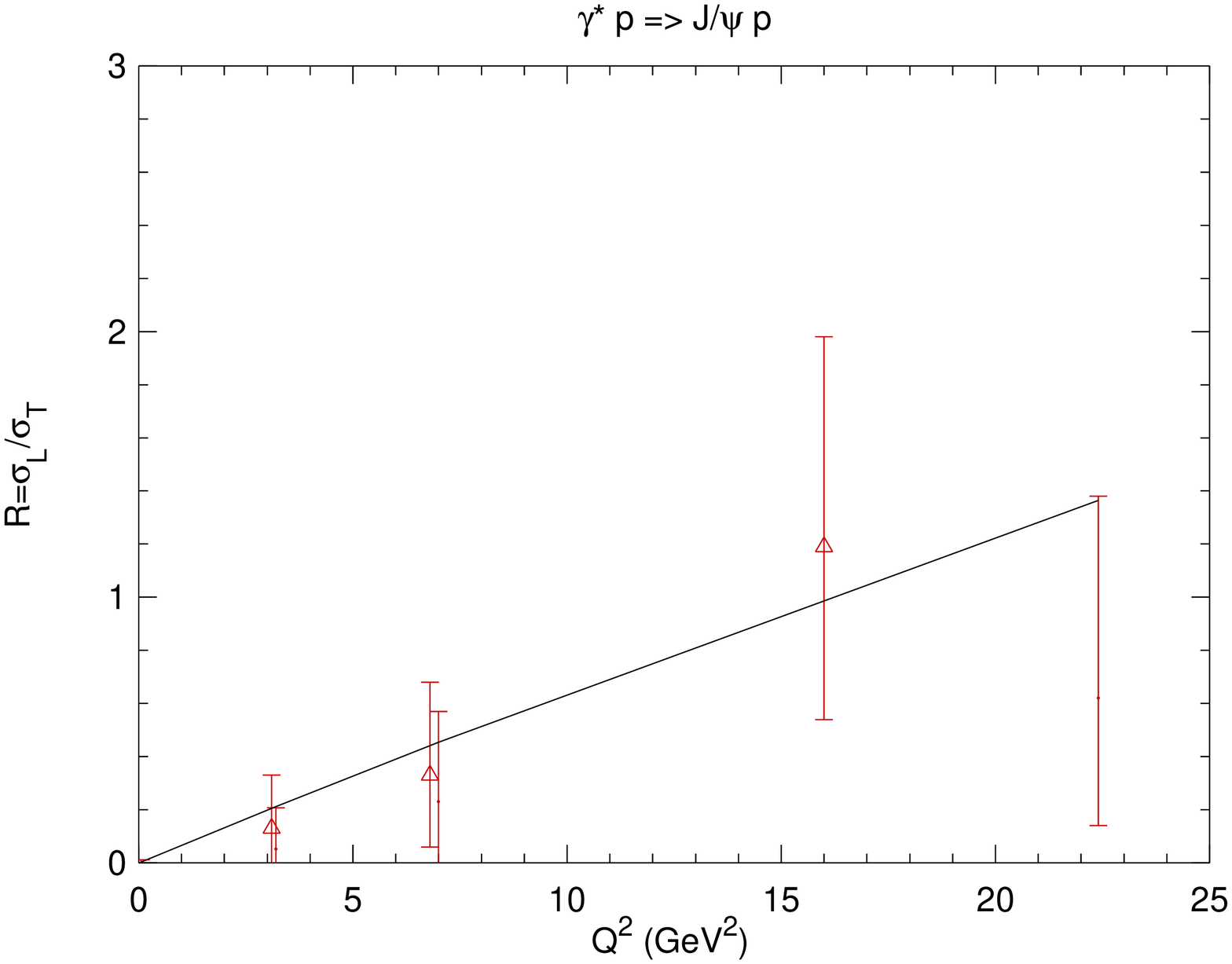}
\caption{The ratio of $R = \frac{\sigma_L}{\sigma_T}$ cross section for longitudinally to transverly polarized vector mesons.  The plots are for $\rho$ (top left $W=75 \; {\rm GeV}$ and right $W=90\; {\rm GeV}$), $\phi$ (bottom  left $W=90 \; {\rm GeV}$), and $J/\Psi$  (bottom right $W=75 \; {\rm GeV}$).}\label{fig:r}
\end{figure}

 The ratio of the longitudinal to the transverse part of the cross section $R = \frac{\sigma_L}{\sigma_T}$ was analyzed as well. This ratio  has a significant  sensitivity to the exact form of the wave functions used.  In Fig.~\ref{fig:r}  the calculation is compared with the experimental data as a function of $Q^2$. The data sets shown in figures  are for very wide bins on $W$, and therefore we have shown the curves which correspond to the middle value of the bins. 
 
By inspecting the formulae for the transverse and longitudinal cross sections one would  think that
the ratio should be approximately independent of the energy $W$. This is not entirely true as the longitudinal and transverse components of the cross sections are sensitive to somewhat different distributions of the dipole size configurations in the photon wave functions. It is well known that the transverse photon wave function takes more contributions from the large dipole sizes, and therefore one should expect that the energy dependence of $\sigma_T$ should be flatter than that of $\sigma_L$. Experimental data show however that the ratio $R$ does not depend on the $W$ indicating that the large dipole components may be suppressed in the transversly polarized exclusive vector meson production.   
  The overall description is very good with the exception of the $\phi$ data where the theoretical curves indicate the growth with the energy which is slightly too fast as compared with the data.  This trend is similar  to what was observed in \cite{Kowalski:2006hc} which indicates a generic feature of the calculations based on the dipole model.

%%%%%%%%%%%%%%%%%%%%%%%%%%%%%%%%%%%%%%%%%
%%%%%%%%%%%%%%%%%%%%%%%%%%%%%%%%%%%%%%%%%

%%%%%%%%%%%%%%%%%%%%%%%%%%%%%%%%%%%%%%%%%
\subsection{Differential cross section and $t$-distribution in vector meson production}

\begin{figure}
\centering
\includegraphics[angle=270,width=0.45\textwidth]{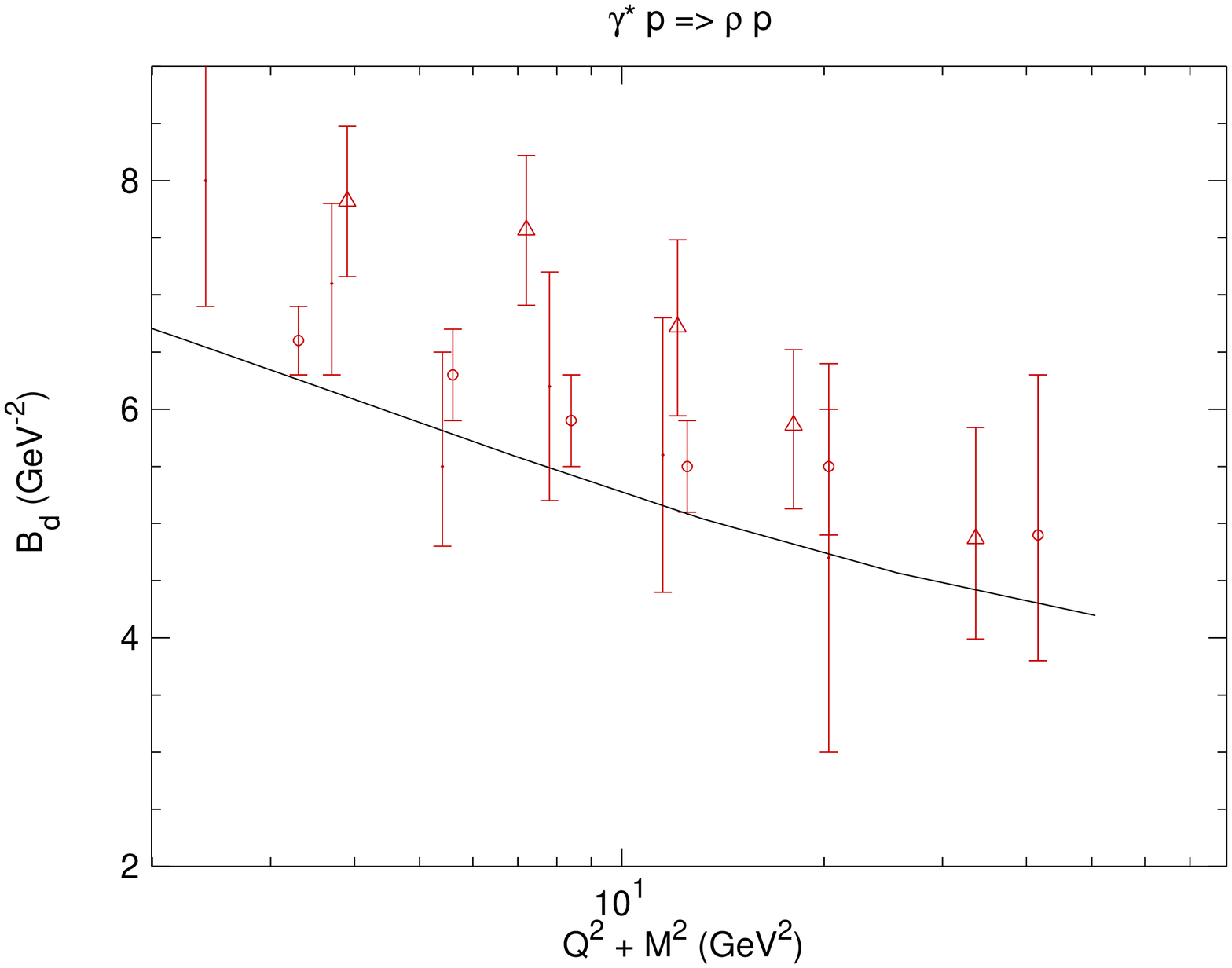}
\includegraphics[angle=270,width=0.45\textwidth]{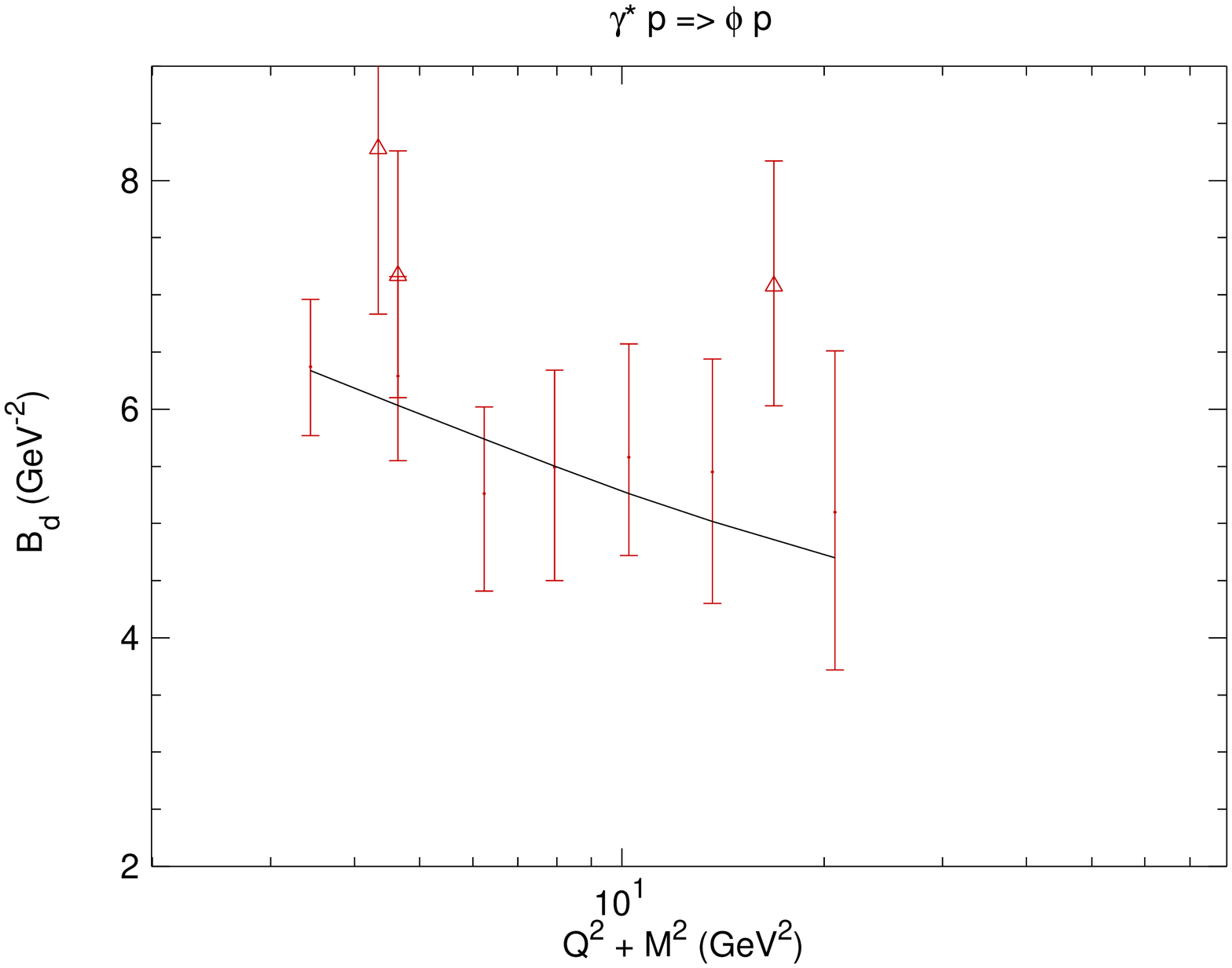}
\includegraphics[angle=270,width=0.45\textwidth]{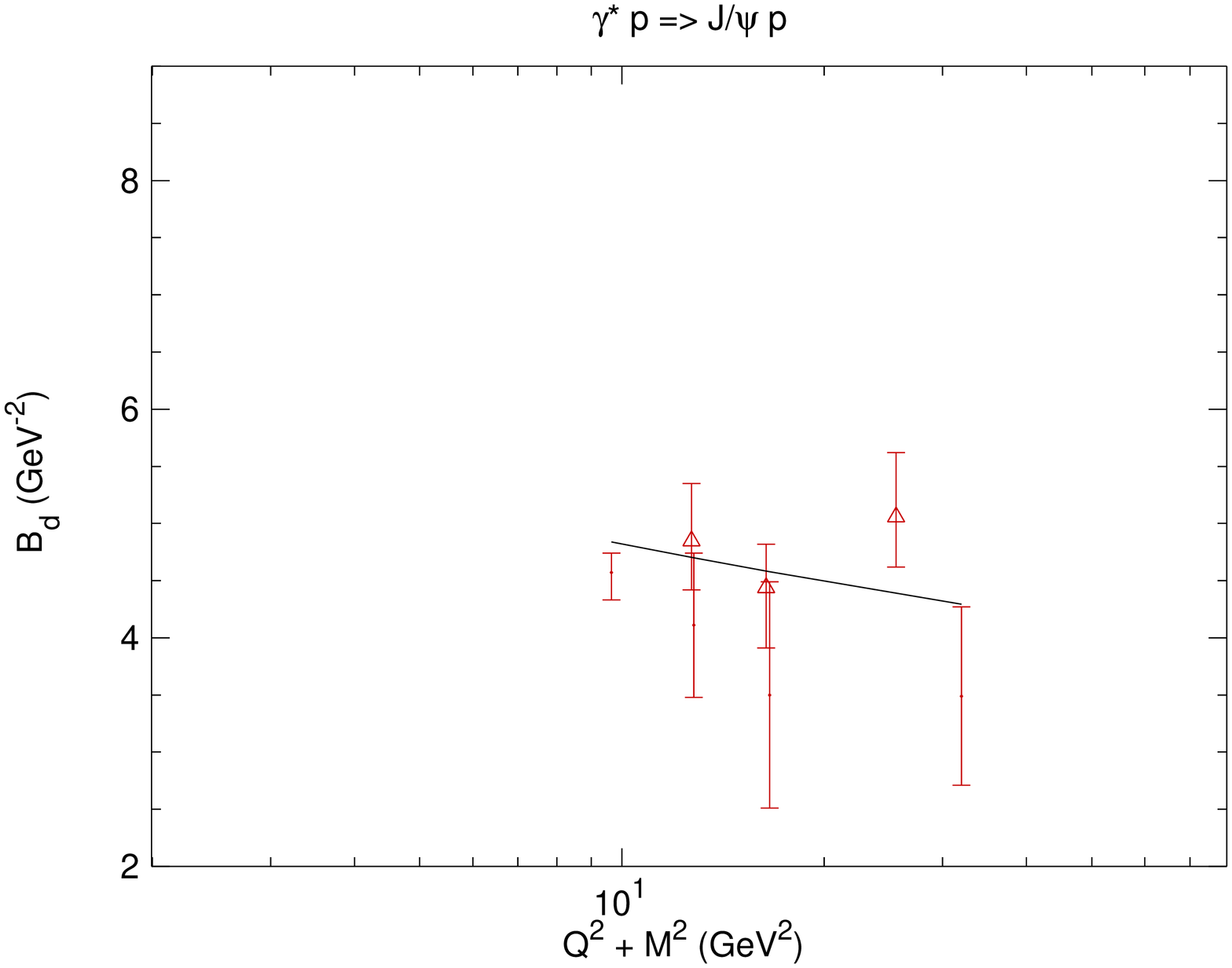}
\caption{Dependence of the slope parameter $B_D$ on combined variable $Q^2+M_V^2$ for $\rho,\phi,J/\Psi$}
\label{fig:BD}
\end{figure}

The $t$-distribution of the differential cross section for elastic vector meson production  is key in unraveling the impact parameter profile of the target at small-$x$.  The $t$-dependence can be related  to the impact parameter dependence via two-dimensional Fourier transform of the amplitude as indicated in Eqs.~(\ref{eq:amplitude}),(\ref{eq:Fourier}).  The differential cross section is usually  parameterized as  $\frac{d\sigma}{dt} \propto e^{-B_D |t|}$ in bins of $Q^2$ and $W$.  The dimensionful slope parameter $B_D$ thus contains the information on the spatial distribution of the interaction region in the scattering process. Three plots in Figure \ref{fig:BD} show the dependence of the slope parameter on the variable $Q^2+M_V^2$ for $\rho,\phi,$ and $J/\Psi$. The theoretical curves  follow the trend of the experimental data. We observe that for the $\rho$ production the dependence of $B_D$ on $Q^2$ is well described but the normalization is underestimated, which is most probably related to the lower normalization for the resulting integrated cross section. The decrease of the slope for low values of $Q^2+M_V^2$ is related to the initial dependence on the size of the vector meson. For larger values of $Q^2$ the dependence flattens to a common value of $B_D \sim 4 \; {\rm GeV}^{-2}$ for each of the vector meson species.  This flattening and universality at large values of $Q^2$ indicates that in this regime the $B_D$ indeed characterizes the size of the proton through the interaction with the small probe which is  high $Q^2$ dipole.
This characteristic size of the gluon density inside the proton $\sqrt{\langle r^2\rangle}\sim0.6\;{\rm fm}$ is markedly smaller than the electromagnetic radius which is of the order $\sim0.8\;{\rm fm}$. This indicates that the gluon distribution differs from the spatial extension of the quarks in the proton.

%%%%%%%%%%%%%%%%%%%%%%%%%%%%%%%%%%%%%%%%%
%%%%%%%%%%%%%%%%%%%%%%%%%%%%%%%%%%%%%%%%%

\begin{figure}
\centering
\includegraphics[angle=270,width=0.49\textwidth]{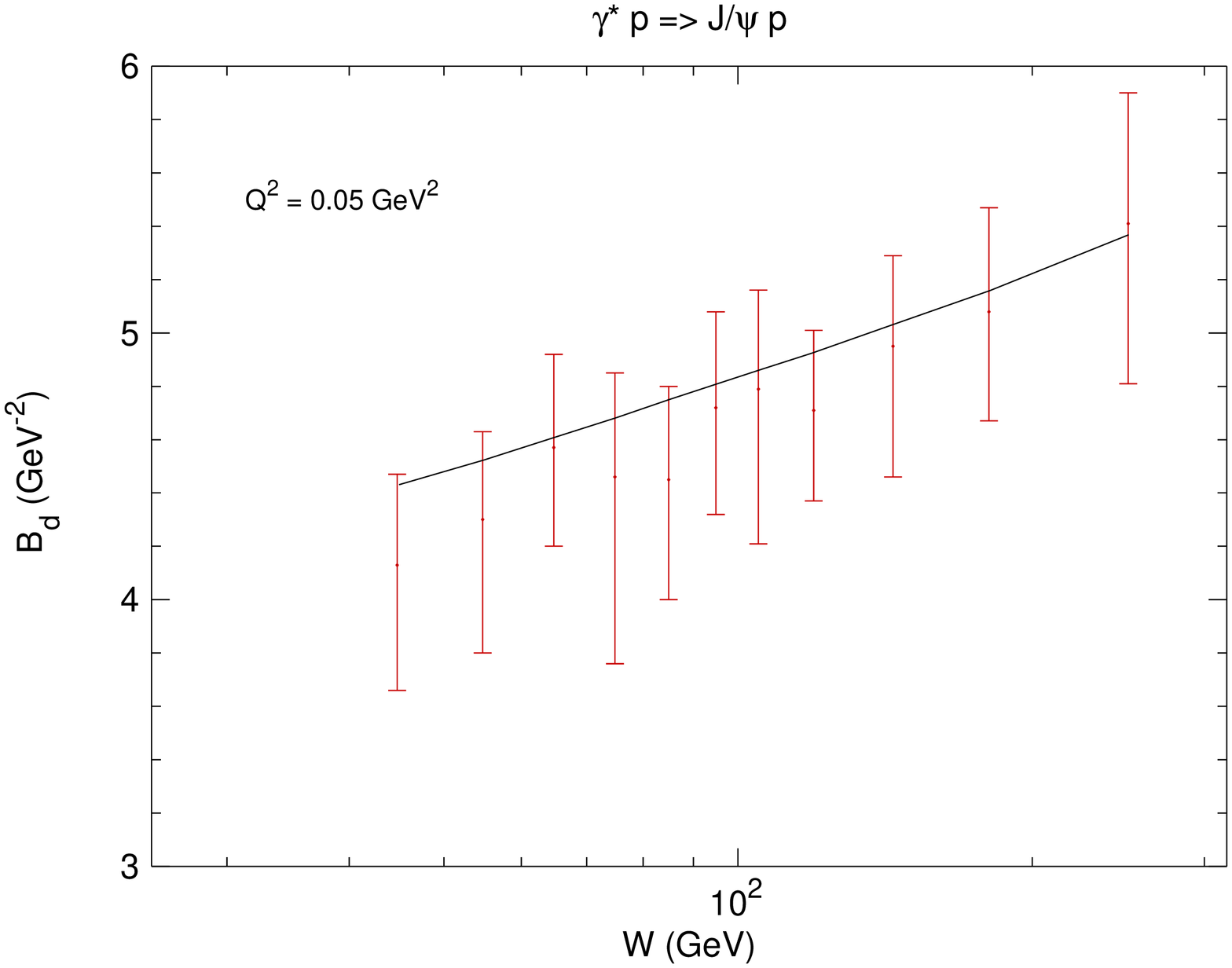}
\includegraphics[angle=270,width=0.49\textwidth]{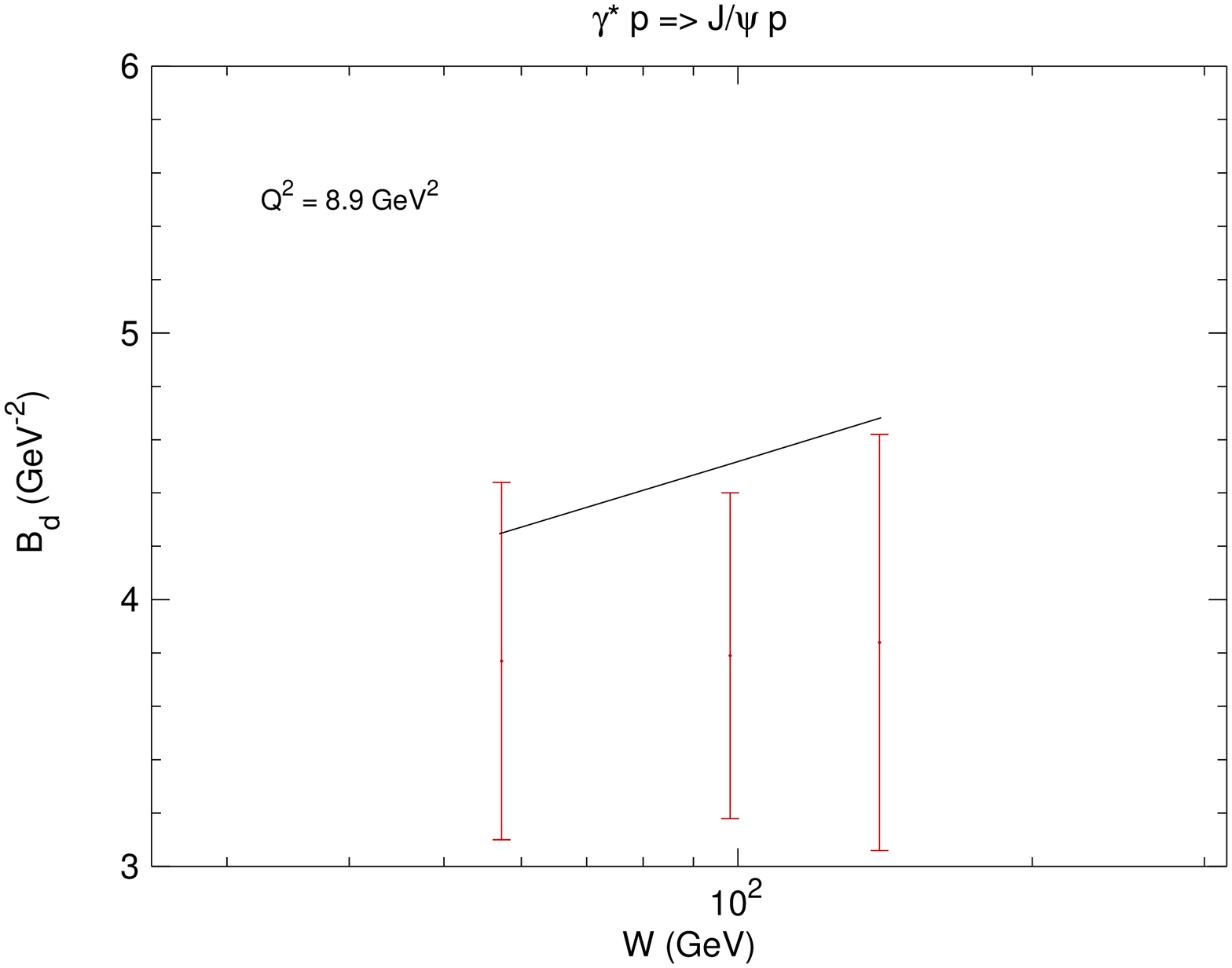}
\includegraphics[angle=270,width=0.49\textwidth]{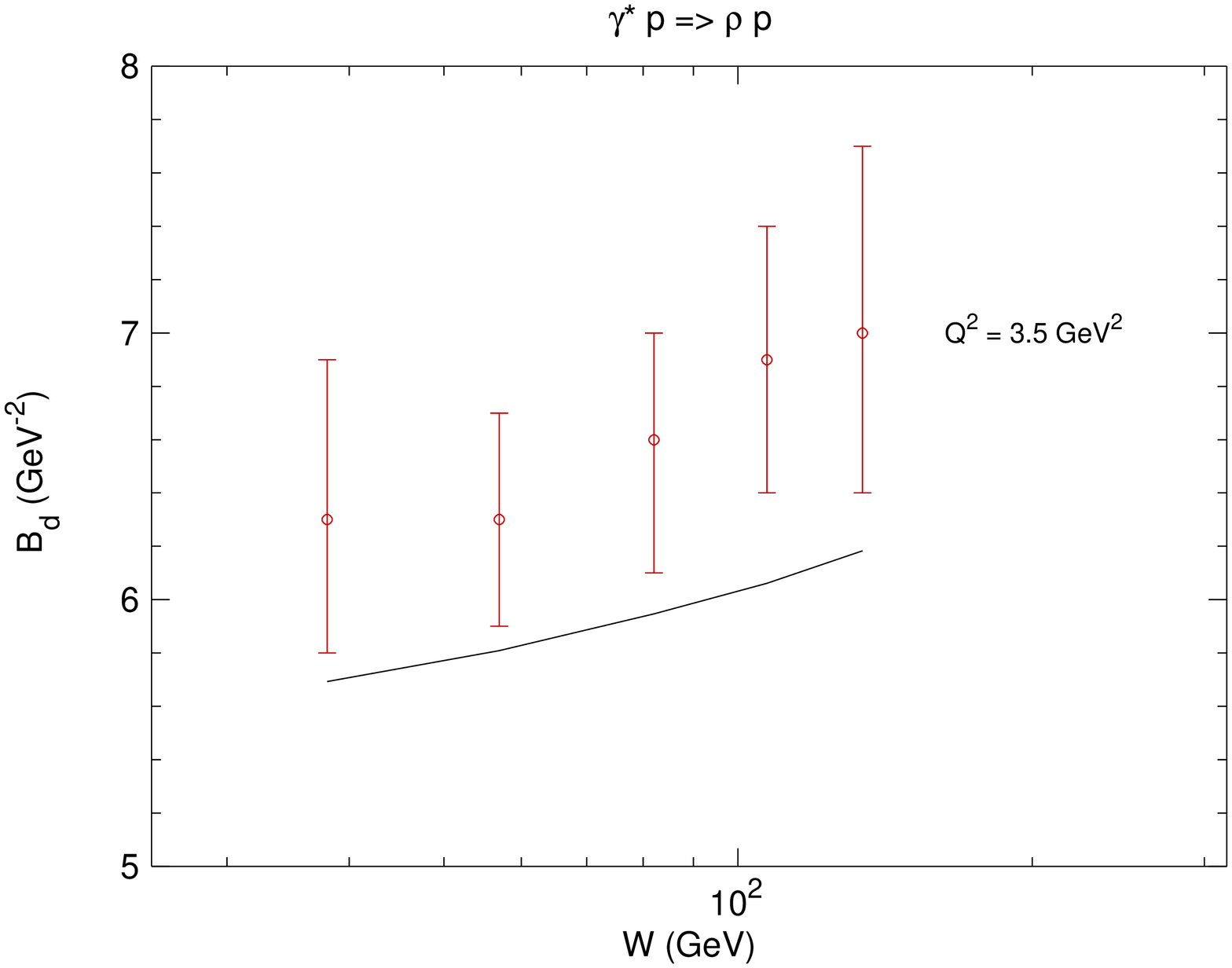}
\caption{Dependence of the slope parameter $B_D$ versus $W$ for $J/\psi$ and $\rho$ production. Data are from H1 experiment  \cite{Aktas:2005xu} and \cite{Chekanov:2007zr}.}
\label{fig:BDW}
\end{figure}

In Fig.~\ref{fig:BDW} we show the same quantity $B_D$ but as a function of $W$ for two different values of $Q^2$ for both $J/\Psi$ and $\rho$. 
  While the error bars on the experimental data for $B_D$  are relatively large, we see that the theoretical curves describe very well the increasing  trend of the data, which is especially  visible in the bin for lower value of $Q^2$ (in fact the bin with higher $Q^2$ is consistent with the flat dependence as well).
  In the calculation presented, the energy dependence of the slope is naturally obtained from the evolution of the dipole scattering amplitude with the energy. The diffusion of the dipoles in the impact parameter space is what provides the change of $B_D$ with energy. Since it is encoded in the BK evolution it naturally leads to the broadening of the impact parameter profile with the energy. 
 The normalization and slope (in energy) of the $B_D$, however, do depend on the two free parameters in the calculation which are not calculated from first principles. The intercept depends on the value of $B_G$ which is used in the initial condition provided by Glauber-Mueller formula, Eqs.~(\ref{eq:glaubermueller}), (\ref{eq:profile}).   The slope in energy of $B_D$ is, on the other hand, controlled by the value of the mass $m$ in the dipole evolution kernel which cuts off the large dipole sizes.
 The $B_D$ parameter for $J/\Psi$ is very well described both in normalization and in the $W$ dependence. On the other hand the 
 the normalization of $\rho$ is once again low, nevertheless the dependence in $W$ is well described. We note that the value of $B_G$ was fitted from the    normalization of the $B_D$ slope of $J/\Psi$ and that the mass $m$ parameter which is related to the $W$ dependence
 is correlated in the calculation with $B_G$. The comparison of the calculation with the data shows that perhaps the initial size in the dipole scattering amplitude may not be universal between $\rho$ and $J/\Psi$ production or that the wave function of $\rho$ needs to include additional non-perturbative components which will increase the interaction size for this meson.
 
%%%%%%%%%%%%%%%%%%%%%%%%%%%%%%%%%%%%%%%%%
%%%%%%%%%%%%%%%%%%%%%%%%%%%%%%%%%%%%%%%%%

\begin{figure}
\centering
\includegraphics[angle=270,width=0.49\textwidth]{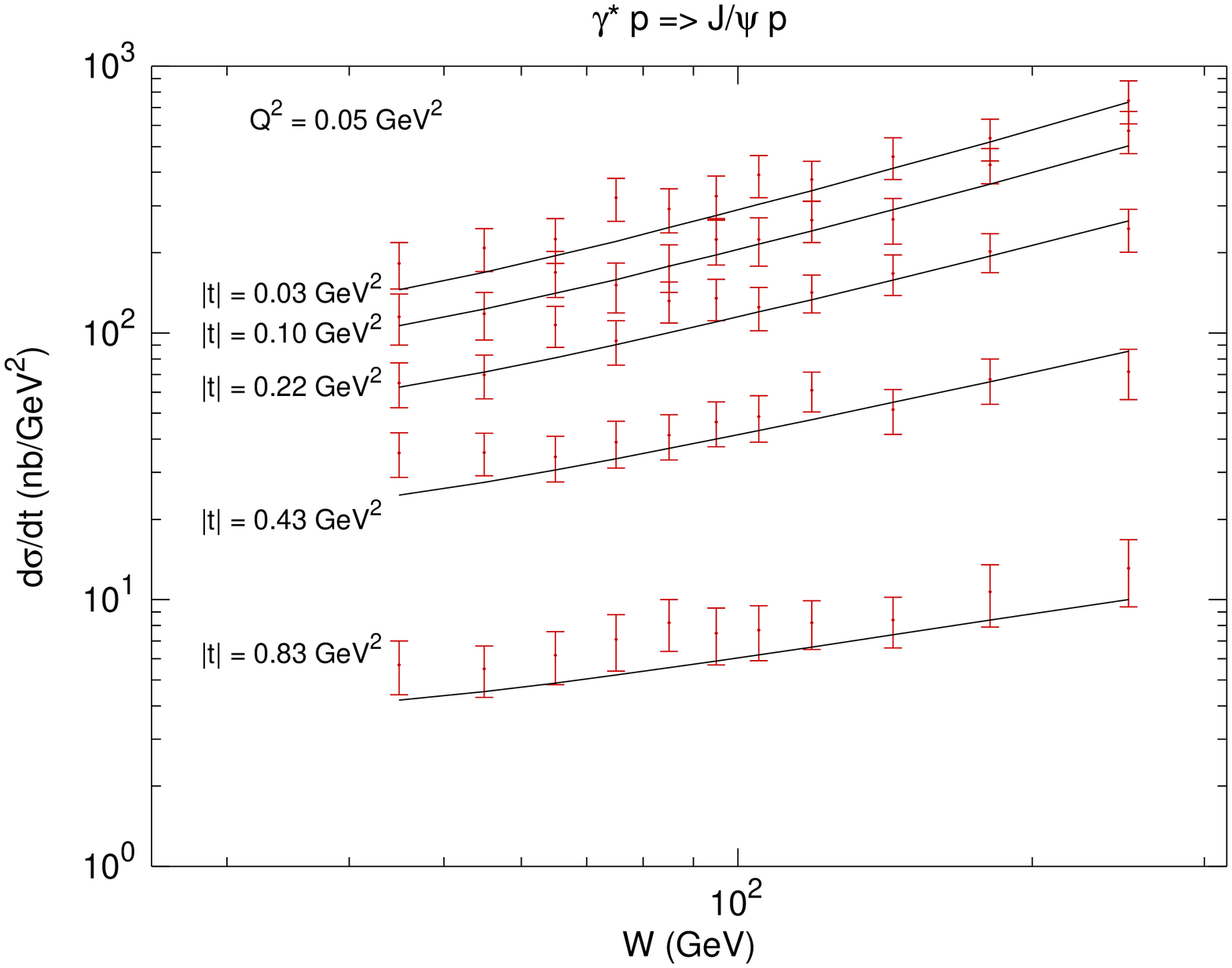}
\includegraphics[angle=270,width=0.49\textwidth]{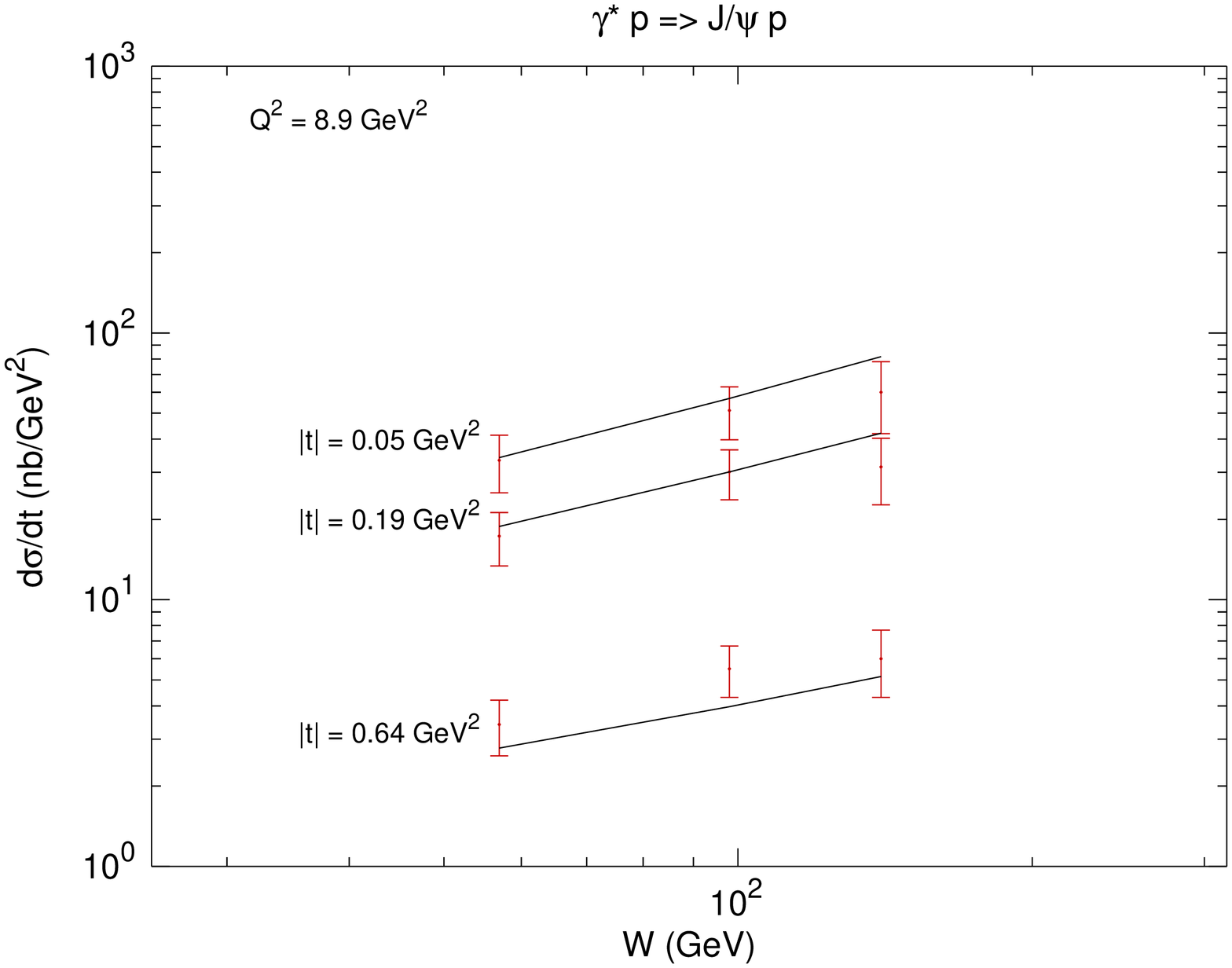}
\caption{The differential cross section of $J/\Psi$ production as a function of $W$ for fixed $Q^2$ in bins of momentum transfer $t$, data from H1 \cite{Aktas:2005xu}. }
\label{fig:dsdtwjpsi}
\end{figure}

\begin{figure}
\centerline{\includegraphics[angle=270,width=0.52\textwidth]{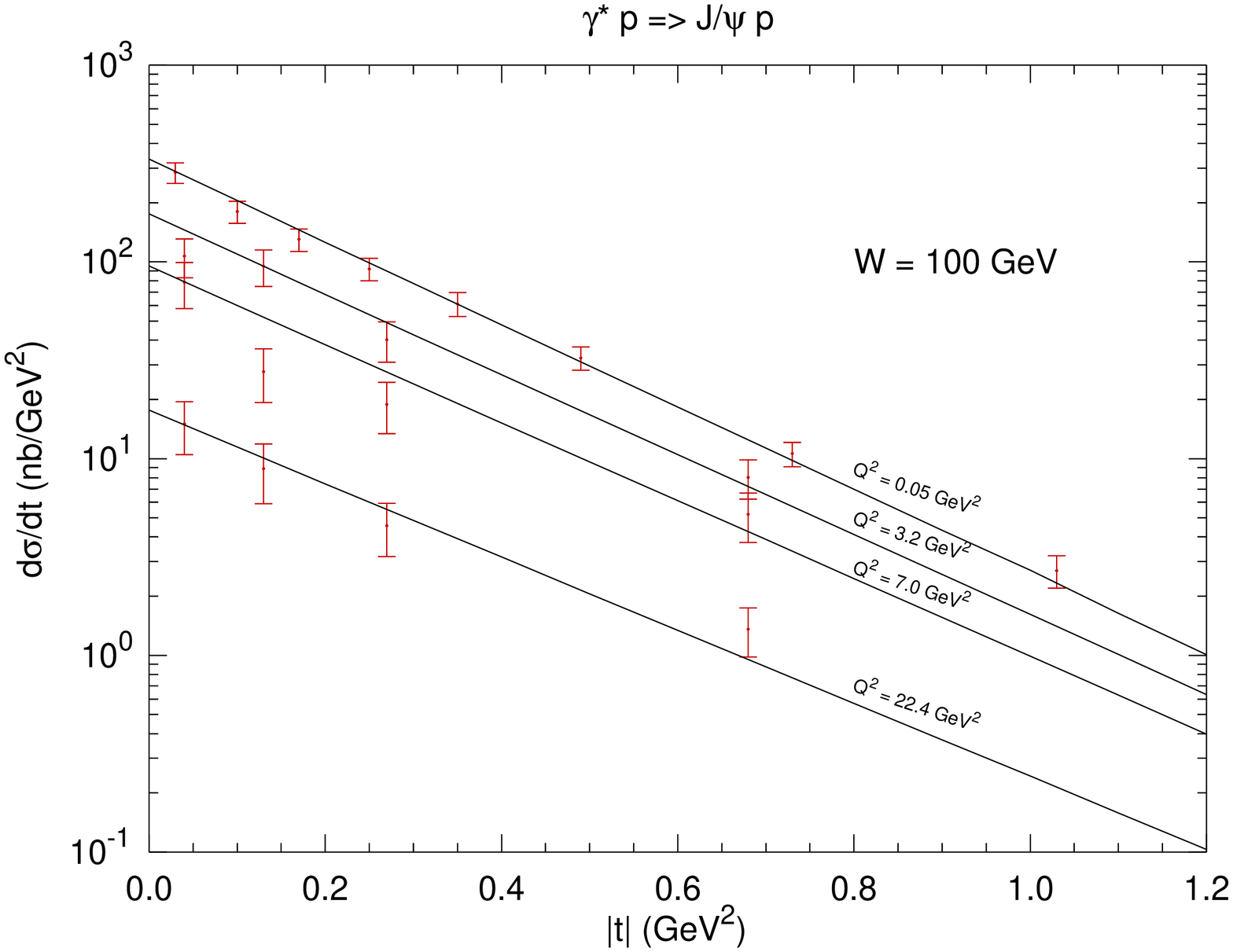}
\includegraphics[angle=270,width=0.52\textwidth]{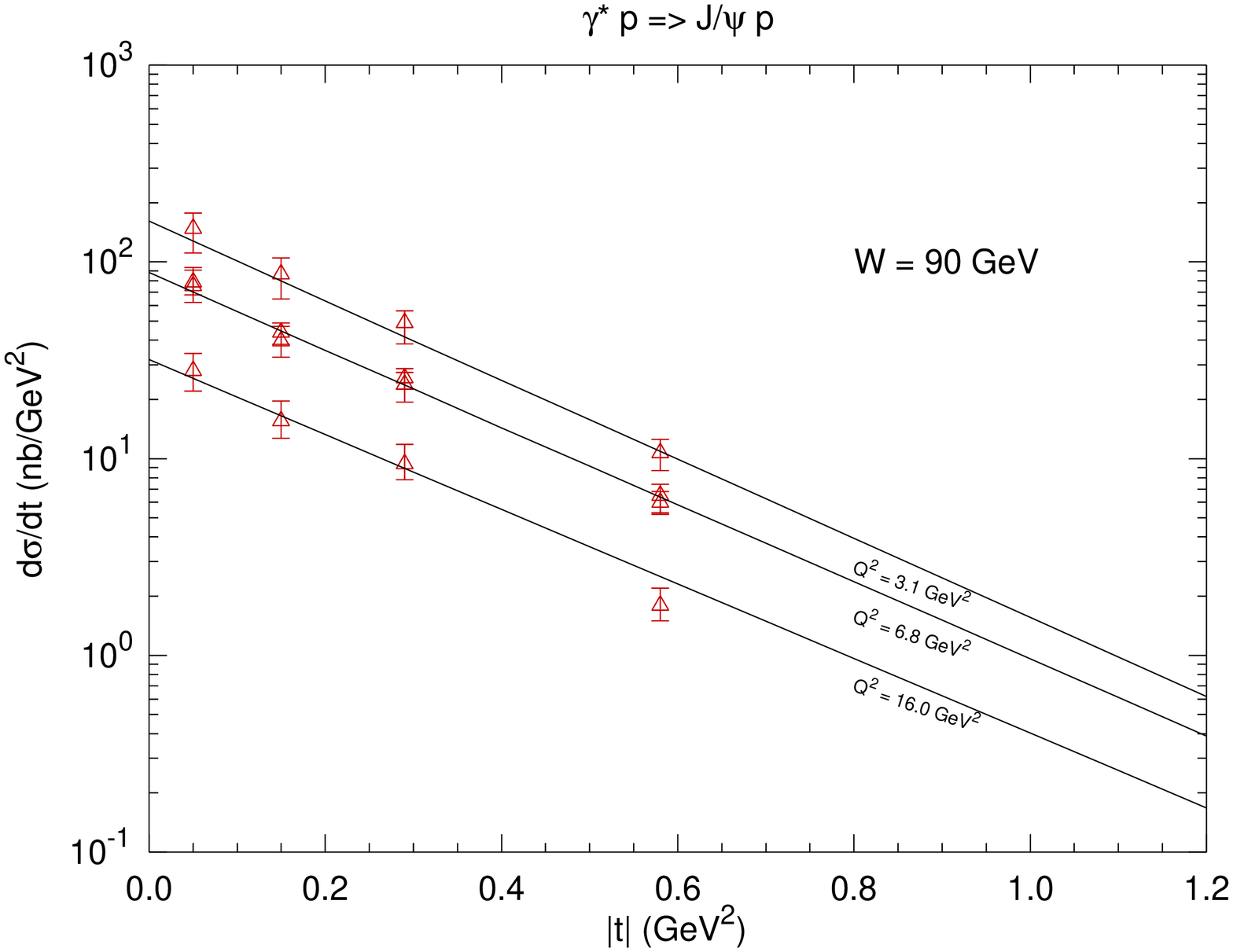}}
\caption{Differential cross section of $J/\Psi$ production for a fixed $W$ in bins of $Q^2$ as a function of momentum transfer $|t|$.   Calculations were done with $W = 100 {\rm GeV}$ and $W=90 {\rm GeV}$. The experimental data  are from H1 experiment \cite{Aktas:2005xu}.}
\label{fig:dsdtjpsi}
\end{figure}

Finally we note that, 
the differential cross section of $J/\Psi$ production compares very favorably with the H1 data \cite{Aktas:2005xu} both in $t$ and in $W$ dependance, which is shown  in Figs.~(\ref{fig:dsdtwjpsi}) and (\ref{fig:dsdtjpsi}).

%%%%%%%%%%%%%%%%%%%%%%%%%%%%%%%%%%%%%%%%%
%%%%%%%%%%%%%%%%%%%%%%%%%%%%%%%%%%%%%%%%%
%%%%%%%%%%%%%%%%%%%%%%%%%%%%%%%%%%%%%%%%%

%%%%%%%%%%%%%%%%%%%%%%%%%%%%%%%%%%%%%%%%%
%%%%Conclusions
\section{Conclusions}
\label{sec:conclusion}

In this paper we have computed the integrated and differential cross section for exclusive diffractive vector meson production in deep inelastic scattering using the dipole model framework for small $x$. The dipole - target scattering amplitude  was obtained from the numerical solution to the impact parameter dependent Balitsky-Kovchegov equation.
We have found that the overall description of the data is very good, meaning that the calculation based on BK equation with impact parameter dependence 
is able to reproduce all of the features seen in the data. Several comments are in order:
\begin{enumerate}
\item For the good description of the experimental data within the framework presented it is important to include
 additional corrections. Non-perturbative modification of the photon wave function at low values of $Q^2$ was necessary, which enhanced the cross sections of the dipoles with sizes of the order of  the hadronic scales.
 \item The skewedness effect was included in the gluon density distribution. This distribution is present in the initial conditions for the small $x$ evolution. This had a substantial impact on the normalization of the resulting cross section and helped to bring the calculations to agreement with the experimental data.
 \item The BK equation was modified to include confinement effects by cutting off large dipole sizes. The parameter $r_{\rm max}=\frac{1}{m}$, which sets the maximal size of the interaction, together with the initial proton size control the slope of the differential cross section with respect to $t$ as well as its variation with the energy. The presented calculation shows very good agreement with the experimental data on $B_D$, including its $W$ dependence in the case of $J/\Psi$.  The slope of $B_D$ is reproduced for $\rho$ but the normalization remains low.  The $W$ dependence is generated dynamically in the dipole evolution. The speed of this increase is controlled  by the parameter $r_{\rm max}=\frac{1}{m}$ which is not calculable from perturbation theory and  needs to be adjusted.
 \item The calculation presented includes the running coupling in the evolution, but misses other important NLL effects which are known to be non-negligible. These should help to bring the calculation to a better agreement with the data, especially as far as the $W$ dependence is concerned. The analysis which includes these effects is thus left for further investigation.
 \end{enumerate}

%%%%%%%%%%%%%%%%%%%%%%%%%%%%%%%%%%%%%%%%%
%%%%Acknowledgments
\section*{Acknowledgments}
 We  thank Henri Kowalski for discussions as well as his assistance by allowing us usage of parts of his fortran code for the evaluation of the initial conditions. We also thank Emil Avsar and Mark Strikman  for discussions. This work was supported  by the   Polish NCN 
grant DEC-2011/01/B/ST2/03915  and the DOE OJI grant No. DE - SC0002145.  A.M.S. is supported by the Sloan Foundation.
%%%%%%%%%%%%%%%%%%%%%%%%%%%%%%%%%%%%%%%%%
%%%%bibliography in BibTex
\bibliographystyle{JHEP}
\bibliography{mybib}

\providecommand{\href}[2]{#2}\begingroup\raggedright\begin{thebibliography}{10}

\bibitem{Adloff:1999kg}
{\bf H1} Collaboration, C.~Adloff {\em et.~al.}, {\it {Elastic
  electroproduction of rho mesons at HERA}},  {\em Eur.Phys.J.} {\bf C13}
  (2000) 371--396, [\href{http://xxx.lanl.gov/abs/hep-ex/9902019}{{\tt
  hep-ex/9902019}}].

\bibitem{Aktas:2005xu}
{\bf H1} Collaboration, A.~Aktas {\em et.~al.}, {\it {Elastic J/psi production
  at HERA}},  {\em Eur.Phys.J.} {\bf C46} (2006) 585--603,
  [\href{http://xxx.lanl.gov/abs/hep-ex/0510016}{{\tt hep-ex/0510016}}].

\bibitem{Aaron:2010}
{\bf H1} Collaboration, F.~Aaron {\em et.~al.}, {\it {Diffractive
  Electroproduction of rho and phi Mesons at HERA}},  {\em JHEP} {\bf 1005}
  (2010) 032, [\href{http://xxx.lanl.gov/abs/0910.5831}{{\tt 0910.5831}}].

\bibitem{Chekanov:2002xi}
{\bf ZEUS} Collaboration, S.~Chekanov {\em et.~al.}, {\it {Exclusive
  photoproduction of J/psi mesons at HERA}},  {\em Eur. Phys. J.} {\bf C24}
  (2002) 345--360, [\href{http://xxx.lanl.gov/abs/hep-ex/0201043}{{\tt
  hep-ex/0201043}}].

\bibitem{Chekanov:2004mw}
{\bf ZEUS} Collaboration, S.~Chekanov {\em et.~al.}, {\it {Exclusive
  electroproduction of J/psi mesons at HERA}},  {\em Nucl.Phys.} {\bf B695}
  (2004) 3--37, [\href{http://xxx.lanl.gov/abs/hep-ex/0404008}{{\tt
  hep-ex/0404008}}].

\bibitem{Chekanov:2005cqa}
{\bf ZEUS} Collaboration, S.~Chekanov {\em et.~al.}, {\it {Exclusive
  electroproduction of phi mesons at HERA}},  {\em Nucl.Phys.} {\bf B718}
  (2005) 3--31, [\href{http://xxx.lanl.gov/abs/hep-ex/0504010}{{\tt
  hep-ex/0504010}}].

\bibitem{Chekanov:2007zr}
{\bf ZEUS} Collaboration, S.~Chekanov {\em et.~al.}, {\it {Exclusive rho0
  production in deep inelastic scattering at HERA}},  {\em PMC Phys.} {\bf A1}
  (2007) 6, [\href{http://xxx.lanl.gov/abs/0708.1478}{{\tt arXiv:0708.1478}}].

\bibitem{Levy:2007fb}
A.~Levy, {\it {Exclusive vector meson electroproduction at HERA}},
  \href{http://xxx.lanl.gov/abs/0711.0737}{{\tt arXiv:0711.0737}}.

\bibitem{Bunyatyan:2008zza}
A.~Bunyatyan, {\it {Exclusive vector mesons and DVCS at HERA}},  {\em
  Nucl.Phys.Proc.Suppl.} {\bf 179-180} (2008) 69--77.

\bibitem{Kowalski:2006hc}
H.~Kowalski, L.~Motyka, and G.~Watt, {\it {Exclusive diffractive processes at
  HERA within the dipole picture}},  {\em Phys.Rev.} {\bf D74} (2006) 074016,
  [\href{http://xxx.lanl.gov/abs/hep-ph/0606272}{{\tt hep-ph/0606272}}].

\bibitem{Frankfurt:2010ea}
L.~Frankfurt, M.~Strikman, and C.~Weiss, {\it {Transverse nucleon structure and
  diagnostics of hard parton-parton processes at LHC}},  {\em Phys.Rev.} {\bf
  D83} (2011) 054012, [\href{http://xxx.lanl.gov/abs/1009.2559}{{\tt
  arXiv:1009.2559}}].

\bibitem{Balitsky:1995ub}
I.~Balitsky, {\it {Operator expansion for high-energy scattering}},  {\em Nucl.
  Phys.} {\bf B463} (1996) 99--160,
  [\href{http://xxx.lanl.gov/abs/hep-ph/9509348}{{\tt hep-ph/9509348}}].

\bibitem{Kovchegov:1999yj}
Y.~V. Kovchegov, {\it {Small-x F2 structure function of a nucleus including
  multiple pomeron exchanges}},  {\em Phys. Rev.} {\bf D60} (1999) 034008,
  [\href{http://xxx.lanl.gov/abs/hep-ph/9901281}{{\tt hep-ph/9901281}}].

\bibitem{Kovchegov:1999ua}
Y.~V. Kovchegov, {\it {Unitarization of the BFKL pomeron on a nucleus}},  {\em
  Phys. Rev.} {\bf D61} (2000) 074018,
  [\href{http://xxx.lanl.gov/abs/hep-ph/9905214}{{\tt hep-ph/9905214}}].

\bibitem{Nemchik:1994fp}
J.~Nemchik, N.~N. Nikolaev, and B.~Zakharov, {\it {Scanning the BFKL pomeron in
  elastic production of vector mesons at HERA}},  {\em Phys.Lett.} {\bf B341}
  (1994) 228--237, [\href{http://xxx.lanl.gov/abs/hep-ph/9405355}{{\tt
  hep-ph/9405355}}].

\bibitem{Frankfurt:1995jw}
L.~Frankfurt, W.~Koepf, and M.~Strikman, {\it {Hard diffractive
  electroproduction of vector mesons in QCD}},  {\em Phys.Rev.} {\bf D54}
  (1996) 3194--3215, [\href{http://xxx.lanl.gov/abs/hep-ph/9509311}{{\tt
  hep-ph/9509311}}].

\bibitem{Nemchik:1996cw}
J.~Nemchik, N.~N. Nikolaev, E.~Predazzi, and B.~Zakharov, {\it {Color dipole
  phenomenology of diffractive electroproduction of light vector mesons at
  HERA}},  {\em Z.Phys.} {\bf C75} (1997) 71--87,
  [\href{http://xxx.lanl.gov/abs/hep-ph/9605231}{{\tt hep-ph/9605231}}].

\bibitem{Munier:2001nr}
S.~Munier, A.~Stasto, and A.~H. Mueller, {\it {Impact parameter dependent S
  matrix for dipole proton scattering from diffractive meson
  electroproduction}},  {\em Nucl.Phys.} {\bf B603} (2001) 427--445,
  [\href{http://xxx.lanl.gov/abs/hep-ph/0102291}{{\tt hep-ph/0102291}}].

\bibitem{Forshaw:2003ki}
J.~R. Forshaw, R.~Sandapen, and G.~Shaw, {\it {Color dipoles and rho, phi
  electroproduction}},  {\em Phys.Rev.} {\bf D69} (2004) 094013,
  [\href{http://xxx.lanl.gov/abs/hep-ph/0312172}{{\tt hep-ph/0312172}}].

\bibitem{Flensburg:2008ag}
C.~Flensburg, G.~Gustafson, and L.~Lonnblad, {\it {Elastic and quasi-elastic $p
  p$ and $\gamma^{*} p$ scattering in the Dipole Model}},  {\em Eur.Phys.J.}
  {\bf C60} (2009) 233--247, [\href{http://xxx.lanl.gov/abs/0807.0325}{{\tt
  arXiv:0807.0325}}].

\bibitem{Avsar:2005iz}
E.~Avsar, G.~Gustafson, and L.~Lonnblad, {\it {Energy conservation and
  saturation in small-x evolution}},  {\em JHEP} {\bf 07} (2005) 062,
  [\href{http://xxx.lanl.gov/abs/hep-ph/0503181}{{\tt hep-ph/0503181}}].

\bibitem{Avsar:2006jy}
E.~Avsar, G.~Gustafson, and L.~Lonnblad, {\it {Small-x dipole evolution beyond
  the large-N(c) limit}},  {\em JHEP} {\bf 01} (2007) 012,
  [\href{http://xxx.lanl.gov/abs/hep-ph/0610157}{{\tt hep-ph/0610157}}].

\bibitem{Avsar:2007xh}
E.~Avsar, {\it {On the Dipole Swing and the Search for Frame Independence in
  the Dipole Model}},  {\em JHEP} {\bf 11} (2007) 027,
  [\href{http://xxx.lanl.gov/abs/0709.1371}{{\tt arXiv:0709.1371}}].

\bibitem{GolecBiernat:2003ym}
K.~J. Golec-Biernat and A.~M. Stasto, {\it {On solutions of the
  Balitsky-Kovchegov equation with impact parameter}},  {\em Nucl. Phys.} {\bf
  B668} (2003) 345--363, [\href{http://xxx.lanl.gov/abs/hep-ph/0306279}{{\tt
  hep-ph/0306279}}].

\bibitem{Berger:2010sh}
J.~Berger and A.~Stasto, {\it {Numerical solution of the nonlinear evolution
  equation at small x with impact parameter and beyond the LL approximation}},
  {\em Phys. Rev.} {\bf D83} (2011) 034015,
  [\href{http://xxx.lanl.gov/abs/1010.0671}{{\tt arXiv:1010.0671}}].

\bibitem{Nikolaev:1990ja}
N.~N. Nikolaev and B.~Zakharov, {\it {Color transparency and scaling properties
  of nuclear shadowing in deep inelastic scattering}},  {\em Z.Phys.} {\bf C49}
  (1991) 607--618.

\bibitem{Nikolaev:1991et}
N.~Nikolaev and B.~G. Zakharov, {\it {Pomeron structure function and
  diffraction dissociation of virtual photons in perturbative QCD}},  {\em
  Z.Phys.} {\bf C53} (1992) 331--346.

\bibitem{Bartels:2003yj}
J.~Bartels, K.~J. Golec-Biernat, and K.~Peters, {\it {On the dipole picture in
  the nonforward direction}},  {\em Acta Phys.Polon.} {\bf B34} (2003)
  3051--3068, [\href{http://xxx.lanl.gov/abs/hep-ph/0301192}{{\tt
  hep-ph/0301192}}].

\bibitem{Kowalski:2003hm}
H.~Kowalski and D.~Teaney, {\it {An Impact parameter dipole saturation model}},
   {\em Phys.Rev.} {\bf D68} (2003) 114005,
  [\href{http://xxx.lanl.gov/abs/hep-ph/0304189}{{\tt hep-ph/0304189}}].

\bibitem{Balitsky:1998ya}
I.~Balitsky, {\it {Factorization and high-energy effective action}},  {\em
  Phys. Rev.} {\bf D60} (1999) 014020,
  [\href{http://xxx.lanl.gov/abs/hep-ph/9812311}{{\tt hep-ph/9812311}}].

\bibitem{Berger:2011ew}
J.~Berger and A.~M. Stasto, {\it {Small x nonlinear evolution with impact
  parameter and the structure function data}},  {\em Phys.Rev.} {\bf D84}
  (2011) 094022, [\href{http://xxx.lanl.gov/abs/1106.5740}{{\tt
  arXiv:1106.5740}}].

\bibitem{Balitsky:2006wa}
I.~Balitsky, {\it {Quark contribution to the small-$x$ evolution of color
  dipole}},  {\em Phys. Rev.} {\bf D75} (2007) 014001,
  [\href{http://xxx.lanl.gov/abs/hep-ph/0609105}{{\tt hep-ph/0609105}}].

\bibitem{Kovchegov:2006vj}
Y.~V. Kovchegov and H.~Weigert, {\it {Triumvirate of running couplings in
  small-x evolution}},  {\em Nucl. Phys.} {\bf A784} (2007) 188--226,
  [\href{http://xxx.lanl.gov/abs/hep-ph/0609090}{{\tt hep-ph/0609090}}].

\bibitem{Martin:1999wb}
A.~D. Martin, M.~Ryskin, and T.~Teubner, {\it {Q**2 dependence of diffractive
  vector meson electroproduction}},  {\em Phys.Rev.} {\bf D62} (2000) 014022,
  [\href{http://xxx.lanl.gov/abs/hep-ph/9912551}{{\tt hep-ph/9912551}}].

\bibitem{Dosch:1996ss}
H.~G. Dosch, T.~Gousset, G.~Kulzinger, and H.~Pirner, {\it {Vector meson
  leptoproduction and nonperturbative gluon fluctuations in QCD}},  {\em
  Phys.Rev.} {\bf D55} (1997) 2602--2615,
  [\href{http://xxx.lanl.gov/abs/hep-ph/9608203}{{\tt hep-ph/9608203}}].

\bibitem{Kulzinger:1998hw}
G.~Kulzinger, H.~G. Dosch, and H.~Pirner, {\it {Diffractive photoproduction and
  leptoproduction of vector mesons rho, rho-prime and rho-prime-prime}},  {\em
  Eur.Phys.J.} {\bf C7} (1999) 73--86,
  [\href{http://xxx.lanl.gov/abs/hep-ph/9806352}{{\tt hep-ph/9806352}}].

\bibitem{Brodsky:1994kf}
S.~J. Brodsky, L.~Frankfurt, J.~Gunion, A.~H. Mueller, and M.~Strikman, {\it
  {Diffractive leptoproduction of vector mesons in QCD}},  {\em Phys.Rev.} {\bf
  D50} (1994) 3134--3144, [\href{http://xxx.lanl.gov/abs/hep-ph/9402283}{{\tt
  hep-ph/9402283}}].

\bibitem{deTeramond:2005su}
G.~F. de~Teramond and S.~J. Brodsky, {\it {Hadronic spectrum of a holographic
  dual of QCD}},  {\em Phys.Rev.Lett.} {\bf 94} (2005) 201601,
  [\href{http://xxx.lanl.gov/abs/hep-th/0501022}{{\tt hep-th/0501022}}].

\end{thebibliography}\endgroup
%%%%%%%%%%%%%%%%%%%%%%%%%%%%%%%%%%%%%%%%%
\end{document}